\documentclass[acmsmall,screen,nonacm]{acmart}

\usepackage{url}
\usepackage{xspace}
\usepackage{booktabs}
\usepackage{makecell}
\usepackage{hyperref}
\usepackage{graphicx}
\usepackage{colortbl}
\usepackage{multirow}
\usepackage{subfigure}
\usepackage{tablefootnote}
\usepackage{color, xcolor}
\usepackage{amsthm,amsmath,amsfonts}
\usepackage{pifont}
\usepackage{enumitem}
\usepackage[capitalise]{cleveref} %

\usepackage{algorithm}
\usepackage{algpseudocode}
\usepackage{algorithmicx}

\usepackage{xspace}
\newcommand{\toolname}{\textsc{Cream}\xspace}
\newcommand{\ie}{\textit{i.e.,}\xspace}
\newcommand{\eg}{\textit{e.g.,}\xspace}

\newcommand{\etal}{\textit{et al.}\xspace}

\newcommand{\query}{$q'$\xspace}

\usepackage[most]{tcolorbox}

\usepackage[normalem]{ulem}

\newcommand{\delete}[1]{}

\AtBeginDocument{
  \providecommand\BibTeX{{
    \normalfont B\kern-0.5em{\scshape i\kern-0.25em b}\kern-0.8em\TeX}}}

\setcopyright{acmcopyright}
\copyrightyear{2024}
\acmYear{2024}
\acmDOI{1}

\acmJournal{TOSEM}
\acmVolume{1}
\acmNumber{1}
\acmArticle{1}
\acmMonth{1}

\begin{document}
\title{No Man is an Island: Towards Fully Automatic Programming by Code Search, Code Generation and Program Repair}

\author{Quanjun Zhang} 
\orcid{0000-0002-2495-3805}
\email{quanjun.zhang@smail.nju.edu.cn}

\author{Chunrong Fang} 
\orcid{0000-0002-9930-7111}
\email{fangchunrong@nju.edu.cn}

\author{Ye Shang} 
\orcid{0000-0002-2495-3805}
\email{201250032@smail.nju.edu.cn}

\author{Tongke Zhang} 
\orcid{0000-0002-2495-3805}
\email{201250032@smail.nju.edu.cn}

\author{Shengcheng Yu} 
\orcid{0000-0002-2495-3805}
\email{201250032@smail.nju.edu.cn}

\author{Zhenyu Chen}
\email{zychen@nju.edu.cn}
\orcid{0000-0002-9592-7022}

\affiliation{
  \institution{State Key Laboratory for Novel Software Technology, Nanjing University}
  \city{Nanjing}
  \state{Jiangsu}
  \country{China}
}

\begin{abstract}

Automatic programming attempts to minimize human intervention in the generation of executable code, and has been a long-standing challenge in the software engineering community.
To advance automatic programming, researchers are focusing on three primary directions:
(1) code search that reuses existing code snippets from external databases;
(2) code generation that produces new code snippets from natural language;
and (3) program repair that refines existing code snippets by fixing detected bugs.
Despite significant advancements, the effectiveness of state-of-the-art techniques is still limited, such as the usability of searched code and the correctness of generated code.

Motivated by the real-world programming process, where developers usually use various external tools to aid their coding processes, such as code search engines and code testing tools,
in this work, we propose \toolname{}, an automatic programming framework that leverages recent large language models (LLMs) to integrate the three research areas to address their inherent limitations.
Our insight is that the integration of three research areas can overcome their inherent limitations:
the code generator can benefit from the valuable information retrieved by the code searcher, while the code repairer can refine the quality of the generated code with external feedback.
In particular, our framework first leverages different code search strategies to retrieve similar code snippets, which are then used to further guide the code generation process of LLMs.
Our framework further validates the quality of generated code by compilers and test cases, and constructs repair prompts to query LLMs for generating correct patches.
We conduct preliminary experiments to demonstrate the potential of our framework, \eg helping CodeLlama solve 267 programming problems with an improvement of 62.53\%.
As a generic framework, \toolname{} can integrate various code search, generation, and repair tools, combining these three research areas together for the first time.
More importantly, it demonstrates the potential of using traditional SE tools to enhance the usability of LLMs in automatic programming.

\end{abstract}



\maketitle

\section{Introduction}
Software automation has long been a vision of \textbf{Software Engineering (SE)}, with one of the significant challenges being the task of automatic programming~\cite{mei2018can,roychoudhury2019automated}.
Automatic programming attempts to handle high-level specifications (\eg natural language and test cases) into correct source code without direct human intervention~\cite{wang2023natural}.
It effectively reduces manual coding effort, improves efficiency, and minimizes programming errors, thus enabling a more robust software development pipeline.
Besides, it democratizes programming by making it accessible to individuals with varying levels of programming expertise.
This is particularly significant as software permeates various industries in modern society, allowing domain-specific experts, who may not be proficient in programming, to undertake programming tasks tailored to their specific needs, such as AI applications in Science~\cite{wang2023scientific}.

In the SE literature, to advance automatic programming effectively, researchers are focusing on three primary directions:

\begin{itemize}
    \item \textbf{Code Search.}
    This task~\cite{sun2024survey,xie2024survey,kim2024big,liu2022opportunities} involves developing sophisticated algorithms to search and retrieve existing code snippets from vast databases or the internet. 
    Code search is able to accelerate development and promote best practices by enabling the reuse of code.
    
    \item \textbf{Code Generation.}
    This task~\cite{zan2023large,yu2024codereval,du2023classeval,mastropaolo2023robustness,yan2023closer,wei2023towards,zhu2024hot} explores the automatic creation of code based on high-level specifications or requirements. It harnesses advanced machine learning models and artificial intelligence to translate human intentions into functional programs.
    
    \item \textbf{Program Repair.}~\cite{gazzola2019automatic,goues2019automated,monperrus2019automatic,zhang2024survey,zhang2024systematic}
    This task involves fixing bugs or vulnerabilities in existing code during the software maintenance phase.
    Program repair can be seen as automatic code generation at a micro-scale by generating correct code from buggy code.
    
\end{itemize}

Over the last decade, considerable research efforts have been devoted to advancing the state-of-the-art in the three areas.
Despite being promising, existing research in these domains still suffers from several limitations.
First, prior studies~\cite{sun2024survey,di2023code} find that code snippets retrieved by code search techniques cannot be directly reused and require manual adaptation, consuming a significant amount of time.
Second, code generation techniques often struggle to produce syntactically and semantically correct code that can pass both the compiler and test cases.
Recent studies~\cite{fan2023automated,liu2024refining} show that even the latest \textbf{Large Language Models (LLMs)} still tend to generate code that contains errors and vulnerabilities.
Third, research on program repair~\cite{zhang2024survey,zhang2024systematic} is mostly confined to semantic bugs introduced by developers, while overlooking the rapidly emerging field of auto-generated code~\cite{zhang2023autocoderover}.
Therefore, addressing the aforementioned issues can help enhance the effectiveness and usability of code search, code generation, and program repair tools when applied in real-world automatic programming scenarios.

To that end, our insights come from the limitations of existing techniques:

\begin{itemize}
    \item \textbf{Limitation of Code Search.}
Code search is an effective method for finding usable code from external codebases.
However, the retrieved code typically cannot be deployed directly due to several reasons, such as project context, software bugs, and library dependencies~\cite{sun2024survey}.
Thus, developers need to adapt retrieved code to specific requirements, including adjusting variable names, optimizing performance or efficiency, fixing bugs or securities, and including necessary dependencies.
When applied in practice, although code search tools can significantly accelerate development by providing useful code snippets, developers will often expend considerable effort to customize and validate the retrieved code before integrating it into their projects.
To address this issue, a feasible direction is to refine the retrieved code to meet certain requirements automatically.
In this regard, \textbf{program repair is promising to adapt the code retrieved by code search tools with minor modifications}, as the retrieved code may already be very similar to the desired output.

\item \textbf{Limitation of Code Generation.}
Code generation is the focus of LLMs in the SE community, and has achieved continuous progress.
However, LLMs are trained on vast datasets up to a certain cutoff point, making it difficult to acquire and update up-to-date knowledge.
Although fine-tuning remains a possible solution, it is impractical to frequently update LLMs with the latest information due to the vast number of model parameters and computational resources.
Thus, when generating code, LLMs may suffer from outdated knowledge and project-specific context.
Particularly, LLMs are unaware of new knowledge (such as libraries or frameworks) after the last training update, and fail to incorporate information about project-specific requirements, dependencies, or evolving codebases, thus limiting the effectiveness of code generation.
To address this issue, a viable approach is to dynamically search for valuable information to augment the code generator.
\textbf{In this regard, code search can provide useful hints, which can guide LLMs to avoid invalid results during code generation.}

\item \textbf{Limitation of Program Repair.}
As a crucial phase of automatic programming, program repair has achieved significant progress in terms of the number of correctly-fixed bugs~\cite{zhang2023survey}.
However, existing repair research is mainly limited to fixing bugs detected by functional test cases from well-constructed benchmarks~\cite{just2014defects4j,zhang2023critical}.
Recently, LLMs have demonstrated impressive capabilities in automatically generating source code. 
However, the reliability and quality of auto-generated code are usually imperfect~\cite{liu2024refining}, making it difficult to deploy such code directly into projects.
In fact, LLMs may generate source code with syntax and semantic errors as the generation process is static without external validation tools, such as compilers.
This concern raises a significant question: \textit{can we automatically refine the code generated by LLMs to make it sufficiently trusted for integration into software systems}?
Therefore, combining test-driven repair with LLMs can provide dynamic feedback, allowing LLMs to iteratively generate accurate code.
\textbf{In this regard, program repair is promising to help LLMs perform self-debugging with test-driven feedback during the code generation phase.}

\end{itemize}

\begin{figure}
    \centering
    \includegraphics[width=0.99\linewidth]{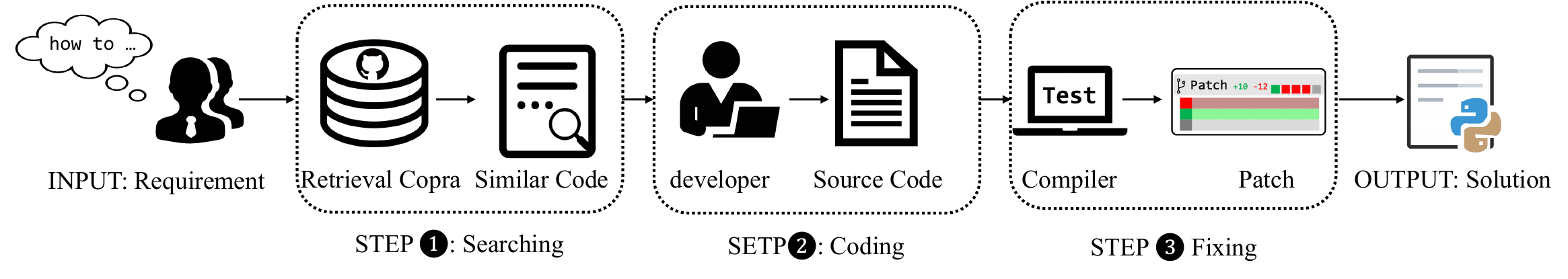}
    \caption{A common programming scenario during software development}
    \label{fig:programming}
\end{figure}

Our analysis motivates us to leverage the complementary strengths of code search, code generation, and program repair techniques to achieve mutual improvement.
In real-world programming scenarios, as shown in Fig.~\ref{fig:programming}, developers typically follow a three-step process.
First, they construct an appropriate search query (such as natural language descriptions) and use search engines (such as GitHub) to find similar code.
Second, they generate their own code by imitating the retrieved code instead of coding from scratch.
Third, they validate the generated code through a compiler to ensure it meets specifications, such as test cases.
Taking inspiration from the developer practice, we can integrate the three research areas into the programming process, \ie code search retrieves similar code for code generation, and program repair provides dynamic feedback to refine the generated code.
However, the key challenge lies in emulating the human developer role to seamlessly connect the three steps, \ie how to integrate retrieved code for generation and how to refine generated code based on test feedback.
Fortunately, thanks to the powerful natural language and programming language understanding capabilities of LLMs, we can fully automate this process through prompt engineering.
Prior studies demonstrate that LLMs can perform code-related tasks in a manner similar to human conversation, and thus, we are motivated to leverage such capabilities of LLMs to connect code search, code generation, and program repair in a unified programming pipeline.

\textbf{This Work.}
We propose a framework \toolname{}, which leverages code sear\textbf{C}h, code gene\textbf{R}ation, and program rep\textbf{A}ir to push forward the boundaries of automatic progr\textbf{AM}ming in the ear of LLMs.
Our work is motivated by the potential to automatically emulate the common developer practice with the help of LLMs' powerful natural language understanding and programming language generation capabilities.
Particularly, given a programming requirement, \toolname{} follows three steps:
(1) code search: an information retrieval (IR) or deep learning (DL)-based technique searches for relevant code from an external database of previous code snippets that may fit the programming context.
(2) code generation: an LLM-based code generator synthesizes a ranked list of code candidates based on both the programming requirement and the retrieved external code knowledge.
(3) program repair: an LLM-based code repairer slightly refines the token sequence of generated code from the previous step by constructing dynamic prompts with test case feedback.
This framework attempts to integrate three well-known research domains that are often developed in isolation, so as to benefit the whole programming pipeline.
More importantly, the integration not only broadens the application scope of these three research areas but also boosts the capabilities of recent LLMs in resolving programming problems effectively.

\textbf{Preliminary Results.}
We conduct a preliminary experiment to evaluate the effectiveness of \toolname{} by implementing it with two retrieval strategies as the code searcher and an open-source LLM CodeLLama with 7 billion parameters as the code generator and program repairer.
The experimental results on the MBPP benchmark demonstrate that 
(1) \toolname{} is able to help CodeLLama in solving programming problems significantly, \eg generating 267 correct solutions with an improvement of 32.18\%;
(2) the three phases positively contribute to the performance of \toolname{}, \eg code search and program repair improves CodeLlama by 24.75\% and 14.85\%, respectively.
To evaluate the potential of \toolname{} in a more real-world programming scenario, we illustrate several case studies from the CoderEval benchmark by utilizing GitHub Search API to search for relevant code and a black-box LLM ChatGPT to generate and repair code.

To sum up, the contributions of this paper are as follows:
\begin{itemize}
  
  \item \textbf{New Dimension.}
  We open a new direction for integrating LLMs with traditional SE research areas for more powerful automatic programming.
  To the best of our knowledge, this is the first work to reveal the potential of LLMs in bridging the gap between three long-standing yet studied-separately research topics, \ie code search, program repair and program repair.

  \item \textbf{Novel Framework.}
  We propose {\toolname}, a three-stage automatic programming framework built on top of LLMs with code search, code generation, and program repair. 
  \toolname{} is a conceptually generic framework that can be easily integrated with various LLMs, code search, generation and repair tools.

  \item \textbf{Preliminary Evaluation.}
  We demonstrate \toolname{}'s ability to generate correct solutions for competitive programming problems.
  Besides, we present case studies to indicate the potential of \toolname{} in real-world programming scenarios.
    
\end{itemize}

\section{Background \& Related Work}
\label{sec:bg&mv}

\subsection{Large Language Models}
In recent years, LLMs have attracted increasing attention from the industry and academia for their impressive capability of processing natural and programming languages~\cite{zhang2023survey}. 
LLMs are generally based on the Transformer architecture that has two key components: the encoder and the decoder.
The former encodes the input into a vector representation for the model to understand the input, while the latter transforms the encoded representation to the output sequence.
Such LLMs have shown outstanding performance in code-related tasks, such as program repair and code generation~\cite{wang2024software}.
For example, ChatGPT~\cite{chatgpt} and GPT-4~\cite{gpt4} released by OpenAI are known for their ability of conducting dialogues with human beings. 
They can take prompts in natural language and generate relevant code accordingly.
CodeLlama~\cite{codellama} is a family of open-source LLMs specifically trained for source code, and can solve programming problems in a zero-shot situation.
Details about the application of LLMs in SE can be found in recent survey papers~\cite{zhang2023survey,wang2024software}.
However, when programming, such LLMs typically struggle to learn the latest knowledge and interact with external coding tools.
In this work, we aim to improve the programming capabilities of off-the-shelf LLMs by integrating them with code search, code generation, and program repair techniques.

\subsection{Code Search}
A common action that developers take while programming is searching for existing code with similar requirements to reuse.
A variety of code search techniques have been proposed to facilitate the retrieval process, and they can be generally classified into two types: IR-based and DL-based~\cite{sun2024survey}.
IR-based code search techniques usually involve indexing the codebases and using scoring algorithms to calculate the similarity between the query and the target code. 
For example, Lucene~\cite{apache_lucene} is a typical IR-based search engine whose default scoring algorithm is BM25, which considers the word frequency and the lengths of documents to rank candidates in the retrieval corpus.
DL-based techniques leverage deep learning models to encode code snippets into vectors, and retrieve similar code based on the cosine similarity between the vectors.
For example, GraphSearchNet~\cite{liu2023graphsearchnet} is a neural network framework based on bidirectional GGNN to map queries and source code by learning the structural information from them.
For a more comprehensive study on code search, please refer to the work of Sun~\etal~\cite{sun2024survey}.
In this work, we implement \toolname{} with two simple yet effective code searchers, \ie IR-based and DL-based sterategies.

\subsection{Code Generation}

Code generation is a popular task that LLMs are applied to because of its great potential to improve the coding efficiency of developers.
For example, AceCoder \cite{li2024acecoder} retrieves similar code and remove redundant retrieval results to boost the effectiveness of code generation.
SkCoder \cite{li2023skcoder} simulates developers' coding behavior by constructing code sketches from the retrieved similar code and turning the sketch into complete code with an encoder-decoder model.
CodeAgent \cite{tang2024codeagent} proposes a novel repo-level code generation framework that integrates different programming tools including information retrieval tools with the purpose of gathering relevant resources so that LLMs can better understand the problems.
Please refer to the of Jiang~\etal~\cite{jiang2024survey} for a more comprehensive survey.
In this work, we construct prompts augmented by the code searcher to query CodeLlama and ChatGPT as the code generator.  

\subsection{Program Repair}

Program repair aims to automatically fix software bugs, thereby reducing the efforts for manual debugging~\cite{zhang2023survey}. 
Existing repair techniques can be broadly categorized into traditional and learning-based ones. 
Traditional program repair approaches include heuristic-based, constraint-based, and template-based techniques. 
With recent advancements in DL, a variety of learning-based repair approaches have been proposed~\cite{tufano2019empirical,lutellier2020coconut,zhu2021syntax}.
Such learning-based techniques leverage Neural Machine Translation (NMT) models to understand the semantics of the bugs and transform them into the correct code.
For example, CoCoNut~\cite{lutellier2020coconut} utilizes a context-aware NMT architecture to represent the buggy source code and its surrounding context separately.

Recently, LLMs are increasingly being utilized for repair tasks~\cite{zhang2023gamma,zhang2023pre,xia2023automated,jiang2023impact}.
For example, Zhang~\etal~\cite{zhang2023pre} investigate the potential of fine-tuning LLMs in repairing security vulnerabilities.
Xia~\etal~\cite{jiang2023impact} evaluate the fixing capabilities of LLMs for Java single-hunk semantic bugs.
Detailed summarization of program repair studies can be found in recent work~\cite{zhang2024survey,zhang2024systematic}.
However, unlike traditional repair techniques, LLMs' powerful natural language capabilities enable them to incorporate external runtime information, thus facilitating iterative patch generation~\cite{olausson2023self,chen2023teaching}.
In this work, motivated by the self-critical capability of LLMs, we leverage execution feedback to integrate program repair into the programming process.

\section{Framework and Implementation}
\label{sec:approach}

\begin{figure}[t]
    \centering
    \includegraphics[width=0.9\linewidth]{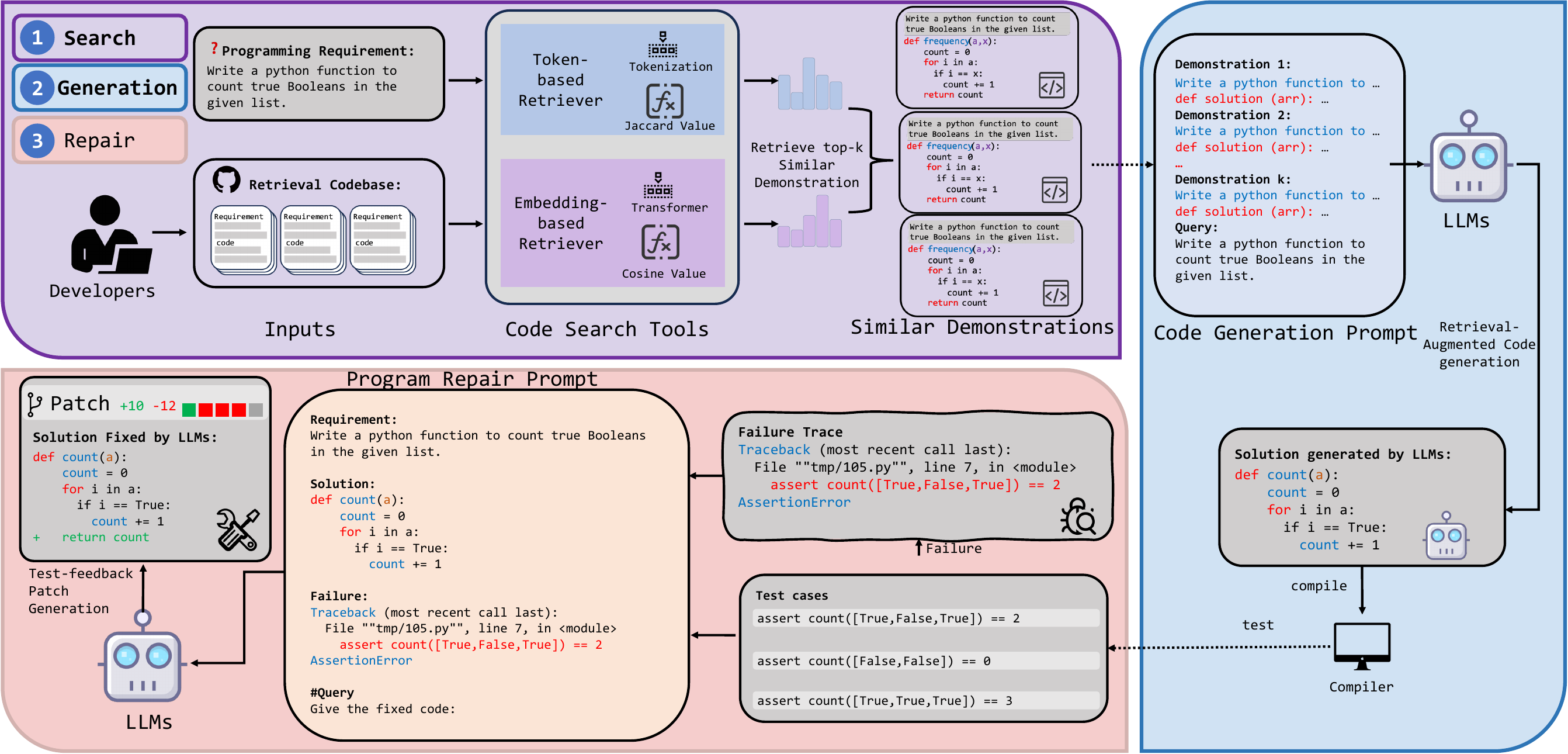}
    \caption{The overall workflow of this paper}
    \label{fig:workflow}
\end{figure}

\subsection{Overview}

\toolname{} takes a programming requirement and external codebases as inputs, and automatically returns source code that meets the requirements.
\cref{fig:workflow} illustrates an overview of \toolname{}, which is divided into three phases: code search, code generation, and program repair.
\begin{itemize}
    \item \textbf{In the code search phase}, given the input as a query, \toolname{} searches a database of source code to find similar code applied previously in a similar context.
    This phase is motivated by the redundancy assumption~\cite{barr2022plastic} that the required code can be found or refactored from other projects with similar contexts.
    In the motivating example in Fig.~\ref{fig:workflow}, given the programming requirement ``\texttt{write a python function to count true Booleans in the given list}'' as query, \toolname{} retrieves a code snippet that addresses a similar programming problem from the codebase: ``\texttt{Write a python function to count true Booleans in the given list}''.
    
    \item \textbf{In the code generation phase}, we first concatenate the retrieved code along with the original programming requirement to construct an augmented input.
    We then query off-the-shelf LLMs, as mentioned in Section~\ref{re:results}, to generate code. 
    Fig.~\ref{fig:workflow} illustrates the generation phase that produces a token sequence for ``\texttt{def count(a): count = 0 for i in a: if i == True: count += 1)}'', which is close to the intended code.
    However, as LLMs generate code tokens from a probabilistic perspective without the ability to dynamically execute and validate the generated code, the generated code may still be imperfect.
    For example, in Fig.~\ref{fig:workflow}, the generated code lacks a return statement, resulting in output values that do not match the expected results and failing to pass the test cases.
    Although the generated code can be quite similar to the intended output, developers still need to spend manual efforts to inspect and modify the generated code.

    \item
    \textbf{In the program repair phase}, we attempt to capture the minor modifications and further refine the generated code with dynamic test feedback.
    Our key insight is that, due to the powerful code generation capabilities of LLMs, the code returned in the last phase is already close to perfect, which can naturally be seen as a program repair task.
    In particular, we utilize the self-critical ability of LLMs by compiling and executing the generated code, then returning the error information to LLMs to guide them in generating more accurate code.
    In the case of Fig~\ref{fig:workflow}, this step explicitly adds the lacked returned statment ``\texttt{return count}'', resulting in the correct patch

\end{itemize}

\subsection{Code Search}
In this phase, given a program specification that needs to be implemented as a query, \toolname{} retrieves relevant code snippets from external databases that are applied before in a similar code context.
\toolname{} utilizes two types of retrieval strategies: an IR-based retriever and a DL-based retriever, to consider the lexical and semantic similarity, respectively.

\subsubsection{IR-based Retriever}
\toolname{} employs the sparse strategy IR as the token-based retriever, to search for a code snippet that is similar to the query programming requirement based on lexical matching.
Suppose $\mathcal{D} = ({r_i, c_i)}^{|\mathcal{D}|}_{i=1}$ be an external retrieval database consisting of $|\mathcal{D}|$ previous code pairs, where $r_i$ is the $i$-th program specification and $c_i$ is its corresponding source code.
Given a query requirement $q$, \toolname{} tokenizes all requirements in the retrieval dataset $\mathcal{D}$ and the query, and removes duplicate tokens for an efficient retrieval process.
\toolname{} then calculates the lexical similarity between $q$ with all requirements in $\mathcal{D}$ based on the Jaccard coefficient.
Jaccard is a widely-used similarity coefficient, to measure the similarity between two sparse vector representations based on their overlapping and unique tokens.
Formula~\ref{equation:jaccard} defines the calculation of Jaccard similarity, where $S(q)$ and $S(r_i)$ are the sets of code tokens of two requirements $q$ and $r_i$, respectively.
The value varies from 0\% to 100\%, and a higher value indicates a higher similarity.
In this work, considering that all requirements are natural language descriptions, we adopt space to tokenize each requirement instead of a sub-word level tokenizer.

\begin{equation}
\small
    Jaccard(q, r_i) = \frac{|S(q) \cap S(r_i) |}{|S(q) \cup S(r_i) |}
\label{equation:jaccard}
\end{equation}

\subsubsection{DL-based Retriever}
\toolname{} employs the pre-trained CodeBERT~\cite{2020fengcodebert} as the embedding-based retriever to search for similar code snippets based on semantic similarity.
In particular, \toolname{} first splits all requirements in $\mathcal{D}$ and the query $q$ into a list of tokens and exploits CodeBERT to transform the tokenized tokens into vector representations.
CodeBERT \toolname{} prepends a special token of [CLS] into its tokenized sequence, and calculates the final layer hidden state of the [CLS] token as the contextual embedding.
{\toolname} then calculates the Cosine similarity between the embeddings of two requirements to measure their semantic relevance.
Cosine similarity is widely adopted in previous studies to measure the semantic relevance of two dense vectors.
Given two vectors, Cosine similarity is calculated based on the cosine of the angle between them, which is the dot product of the vectors divided by the product of their lengths.
Formula~\ref{equation:cosine} defines the calculation of Cosine similarity, where $E(q)$ and $E(r_i)$ denote the embeddings of two requirements $q$ and $r_i$.
\begin{equation}
\small
    Cosine(q, r_i) = \frac{E(q) \cdot E(r_i)}{\| E(q)\|  \|E(r_i) \|}
\label{equation:cosine}
\end{equation}

\begin{algorithm}[t]
\caption{Pseudo Code of Code Search in \toolname{}}
\label{alg:search}
\begin{algorithmic}[1]
    \renewcommand{\algorithmicrequire}{\textbf{Input:}}
    \renewcommand{\algorithmicensure}{\textbf{Output:}}
    \Require $q$: a query as a program specification that needs to be implemented
    \Require $\mathcal{D}$: an external retrieval database for code search
    \Require $k$: the number of similar code snippets to be retrieved
    \Ensure $retrievedResult$: retrieved most similar codes to \query{} from $\mathcal{D}$
    
    \State $similarityList \leftarrow []$
    \label{line:init}
    \For{each code requirement $r_i$ in  $\mathcal{D}$}
        \State $similarityScore \leftarrow \textsc{calculateSimilarity}(q, r_i)$
        \label{line:calcsim}
        \State $similarityList.append(\{requirement:r_i, code:c_i, score:similarityScore\})$
        \label{line:append}
    \EndFor

    \State $sortedList \leftarrow \textsc{customSort}(similarityList)$
    \label{line:sort}
    \State $retrievedResult \leftarrow [c \text{ for } \{requirement, c, score\} \text{ in } sortedList[:k]]$
    \label{line:extact}
    \State \Return $retrievedResult$
    \label{line:return}

    \Function{\textsc{calculateSimilarity}}{$q, r_i$}
    \label{line:calc_start}
        \Comment{Calculate similarity score between $q$ and $r_i$}
        \State $\mathcal{T}(q) \leftarrow \textsc{extractToken}(q)$ \textcolor{blue}{OR} 
        \Statex \hspace{1.24em} \textcolor{blue}{$\mathcal{E}(q') \leftarrow \textsc{extractEmbedding}(q')$}
        
        \State $\mathcal{T}(r_i) \leftarrow \textsc{extractToken}(r_i)$ \textcolor{blue}{OR}
        \Statex \hspace{1.24em} \textcolor{blue}{$\mathcal{E}(r_i) \leftarrow \textsc{extractEmbedding}(r_i)$}
        
        \State $similarityScore \leftarrow \textsc{computeJaccardSimilarity} (\mathcal{T}(q), \mathcal{T}(r_i))$ \textcolor{blue}{OR}
        \Statex \hspace{1.24em} \textcolor{blue}{$similarityScore \leftarrow \textsc{computeCosineSimilarity}(\mathcal{E}(q), \mathcal{E}(r_i))$}
        \State \Return $similarityScore$
    \EndFunction
    \label{line:calc_end}

    \Function{\textsc{customSort}}{$similarityList$}
    \label{line:sort_start}
        \Comment{Sort list of tuples $similarityList$ in descending order by similarity score}
        \For{$i \leftarrow 0$ to $length(similarityList) - 1$}
            \For{$j \leftarrow 0$ to $length(similarityList) - i - 1$}
                \If{$similarityList[j][score] < similarityList[j+1][score]$}
                    \State $temp \leftarrow similarityList[j]$
                    \State $similarityList[j] \leftarrow similarityList[j+1]$
                    \State $similarityList[j+1] \leftarrow temp$
                \EndIf
            \EndFor
        \EndFor
        \State \Return $similarityList$
    \EndFunction
    \label{line:sort_end}
\end{algorithmic}
\end{algorithm}

Algorithm~\ref{alg:search} presents the detailed workflow of the search strategy in our work.
The algorithm starts by taking three inputs: a query representing the program specification to be implemented (\query), an external database containing code snippets ($\mathcal{D}$), and the number of similar code snippets to be retrieved (top-$k$).
The algorithm initializes an empty list called ``similarityList'' to store the similarity scores between the query and each code requirement in the database (Line~\ref{line:init}). 
For each code requirement in the database, the algorithm calculates the similarity score between the query and the code requirement using a function named \textsc{calculateSimilarity} (Line~\ref{line:calcsim}). 
This function extracts tokens or embeddings from both the query and the code requirement, then calculates the similarity score using either Jaccard similarity for tokens or cosine similarity for embeddings (Line~\ref{line:calc_start}$\sim$\ref{line:calc_end}). 
The resulting similarity score, along with the corresponding code requirement and code snippet, is appended to the ``similarityList'' (Line~\ref{line:append}).
Once all similarity scores are calculated, the algorithm sorts ``similarityList'' in descending order based on the similarity scores using a function named \textsc{customSort} (Line~\ref{line:sort}). 
This function employs a bubble sort algorithm to ensure the list is ordered from the highest to the lowest similarity score (Line~\ref{line:sort_start}$\sim$\ref{line:sort_end}).
After sorting, the algorithm extracts the top-$k$ most similar code snippets from the sorted list and stores them in ``retrievedResult'' (Line~\ref{line:extact}). 
Finally, the algorithm returns ``retrievedResult'' as the output, which contains the code snippets that are most similar to the given query (Line~\ref{line:return}). 
Through this systematic approach, the algorithm effectively retrieves the most relevant code snippets from an external database based on the provided query.

\begin{figure}
    \centering
    \includegraphics[width=0.7\linewidth]{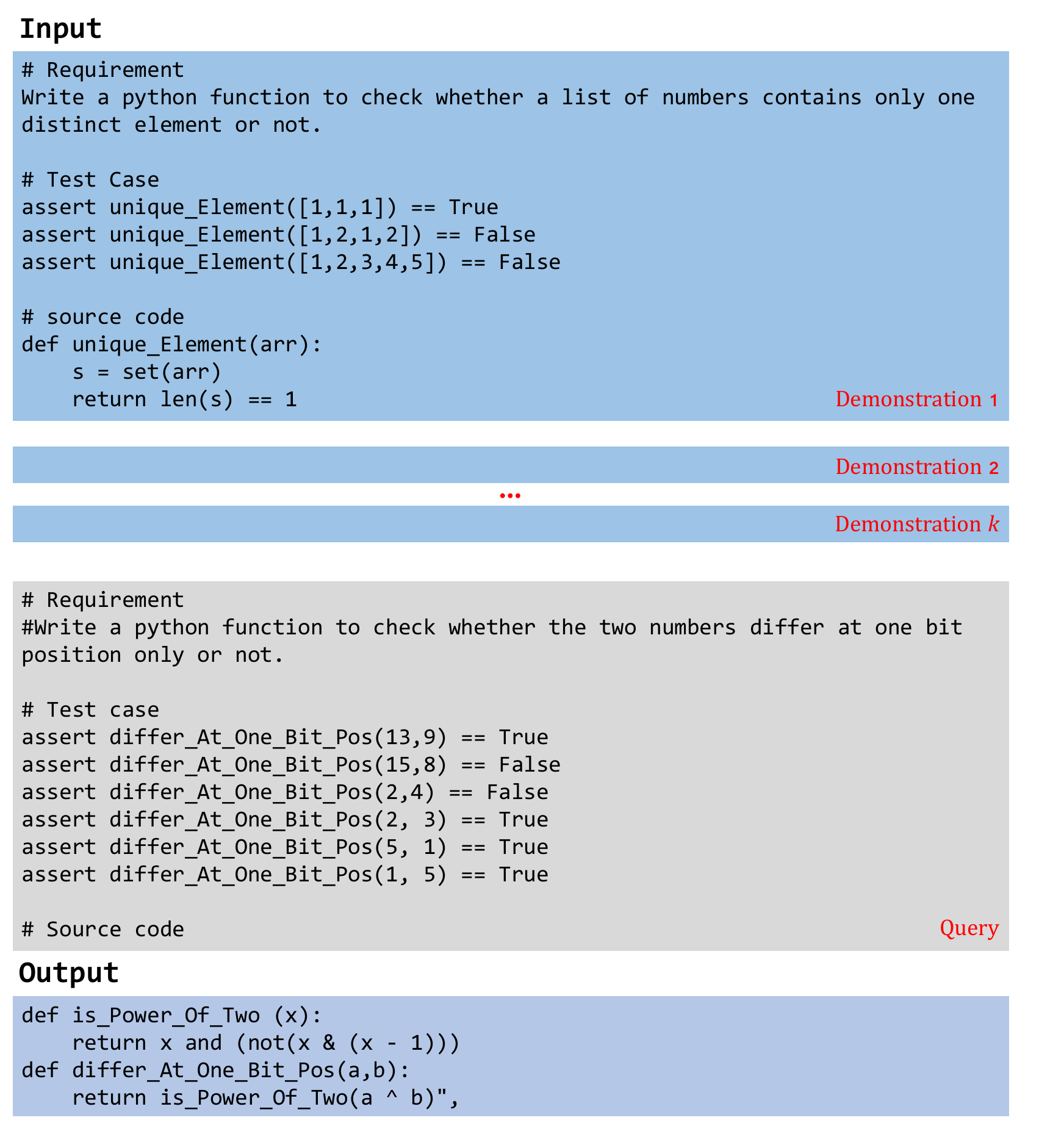}
    \caption{Retrieval-Augmented Code Generation Prompt Template}
    \label{fig:prompt_generation}
\end{figure}

\subsection{Code Generation}

In the code generation phase, we leverage LLMs to generate code based on the programming requirement and the retrieved programming solutions.
Particularly, we build a prompt by composing (a) relevant code demonstrations,  (b) the programming query, and (c) natural language instructions. 
To select demonstrations, we take examples from the demonstration retriever component. 
Then we select a task-specific template and combine these three elements to build the final prompt.

Fig.~\ref{fig:prompt_generation} illustrates a prompt template for code generation. 
The input prompt mainly contains two sections: code demonstrations, and the query.
Each code demonstration section consists of the programming requirement, its test cases, and the expected code. 
The query section contains the natural language instruction beginning with \texttt{\#} in the template, followed by the test cases. 
As it is an autocomplete task, the comment \texttt{Write a python function} is used to signal the model to generate a correct code snippet.
Finally, the expected output for a given prompt is a multi-line code snippet passing the test cases.

\begin{figure}
    \centering
    \includegraphics[width=0.7\linewidth]{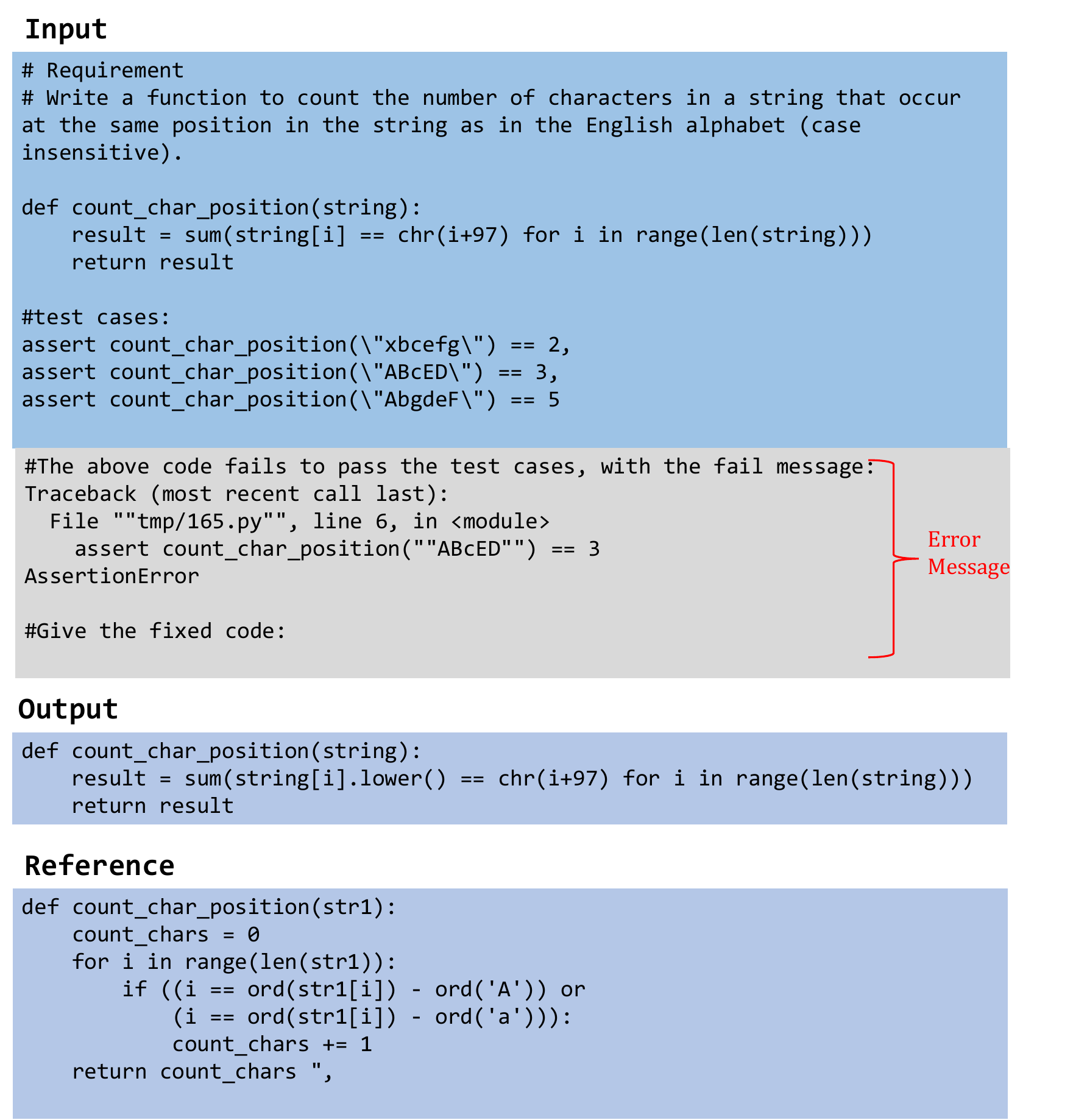}
    \caption{Test-Driven Program Repair Prompt Template}
    \label{fig:prompt_repair}
\end{figure}

\subsection{Program Repair}
\label{sec:approach_repair}
After obtaining generated code snippets by LLMs, this phase attempts to refine them further by fixing syntax and semantic errors.
Particularly, we compile the programs and dynamically execute them against all available test cases.
The test cases provide valuable information about the correctness of the generated code, associated with the error messages if the code does not pass the tests. 
For the failing functions, we input them to the LLMs again together with their requirements and error messages in an attempt to have them fixed. 
To this end, we conduct a dynamic prompt for LLMs in the program repair stage, as shown in Fig.~\ref{fig:prompt_repair}.
The prompt 
The input prompt contains two sections: the programming requirement. its test cases, the generated code, and the failure information.
Each section is denoted by a natural language description that begins with the comment symbol \texttt{\#} in the template.
The query contains the context of test execution followed by the instruction. 
This is essentially an auto-complete task where the query is an incomplete example used to prompt the model. 
Finally, the expected output for a given prompt is a fixed version that addresses the reported failure.

\section{Preliminary Results}
\label{re:results}

We conduct two preliminary experiments to evaluate the performance of our search-generation-repair framework for automatic programming.
We first investigate the performance of \toolname{} in solving competitive programming problems.
We then explore the potential of \toolname{} in solving real-world programming problems.

\subsection{Evaluation 1: The Effectiveness of Three Phases in Solving Competitive Programming Problems}
\label{sec:rq1}

\textbf{Experimental Design}.
In this RQ, we investigate the effectiveness of the code search, code generation, and program repair phases in our automatic programming framework.

We choose Mostly Basic Programming Problems (MBPP) released by Austin~\etal~\cite{austin2021program} as the benchmark. 
It contains 974 easy Python programming problems, their test cases and code solutions obtained from crowdsourcing.
There is also a sanitized version of MBPP with 427 programming problems manually verified and extracted from the full dataset.
We focus on the sanitized MBPP and investigate how effective our framework is in solving problems in it. 
We use the same dataset for code retrieval (except the query) since the coding styles within the same dataset are similar, which helps LLMs better learn what the target code looks like.

We choose CodeLlama as the researched LLM in this RQ and use it to generate Python code for 427 programming problems from the MBPP dataset. CodeLlama is a series of large language models for code generation, and we select the CodeLlama-7b-hf model, which is one of the foundation models with 7B parameters.

We conduct experiments under programming four scenarios to explore how each phase influences the correctness of the generated code:

\begin{itemize}
    \item \textbf{Only code generation.}
    We construct basic prompts only with programming problem descriptions and test cases, and query LLMs to generate Python functions to solve the problems.
    \item \textbf{Code search + code generation.} 
    When using LLMs to generate code, we not only provide basic information of the programming problems, but also additional functions with similar requirements that are retrieved from the same dataset.
    \item \textbf{Code generation + program repair.} 
    We generate code with the basic prompts, and then we use LLMs again to fix the incorrect code it generates.
    \item \textbf{Code search + code generation + program repair (\toolname{}).} 
    We combine all the phases by retrieving similar code, generating code according to the programming requirements as well as the similar code, and repairing the code that does not pass all the tests.
\end{itemize}

\begin{figure}[htbp]
    \centering
    \includegraphics[width=0.6\linewidth]{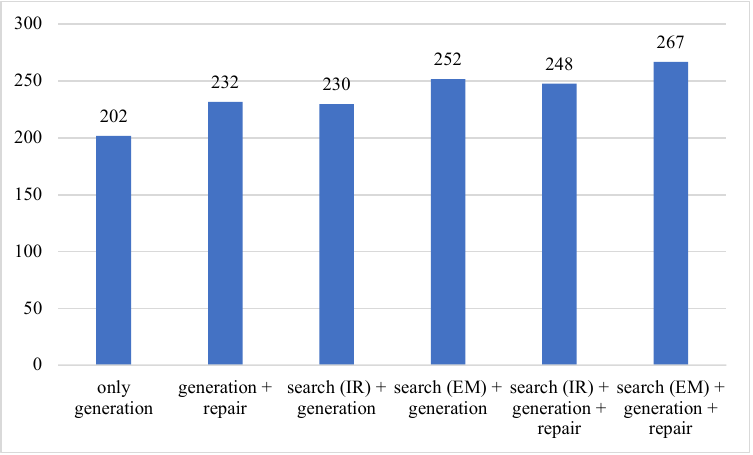}
    \caption{The number of MBPP problems correctly solved by CodeLlama}
    \label{fig:rq1}
\end{figure}
\textbf{Results}.
Fig.~\ref{fig:rq1} presents the number of problems that CodeLlama successfully solves under four different scenarios.
In Fig.~\ref{fig:rq1}, ``search (IR)'' and ``search (EM)'' denote the IR and embedding retrievers, respectively.
When CodeLlama directly performs the code generation task, 202 out of 427 generated solutions are correct, resulting in a correctness rate of 47.31%
Besides, when integrating program repair, code search-IR, and code search-embedding into the programming process, 31, 28, and 50 additional correct solutions are generated, respectively.
Finally, CodeLlama achieves the best performance when code search, code generation and program repair are combined, producing 248 (IR retriever) and 251 (embedding retriever) correct solutions.
The improvement against the code generation-only scenario yields 58.08\% and 62.53\%, respectively.
These results demonstrate that the three-phase pipeline—comprising searching, coding, and repairing—continuously enhances the programming capabilities of LLMs.

\subsection{Evaluation 2: The Potential of Three Phases in Solving Real-world Programming Problems}
\label{sec:rq2}

\textbf{Experimental Design}.
In RQ1, we investigate the performance of \toolname{} in generating functions for competitive programming problems. 
In this RQ, we further explore how effective our approach is in a real-world programming scenario. 

We select CoderEval~\cite{yu2024codereval}. Compared to MBPP and other popular code generation benchmarks which only include standalone functions, CoderEval contains programming tasks extracted from real-world projects as well as separate platforms to execute them, so we can evaluate our automatic programming approach in a real-world scenario.

We simulate real programming scenarios by selecting three functions that need to be implemented from the dataset. First, we use the GitHub search engine to find similar code, then call ChatGPT to generate the code, providing feedback on the test results for corrections.
In the following, we present three real-world examples to illustrate the search-generation-repair capabilities of \toolname{}.
For all three examples, ChatGPT fails to directly generate correct code based solely on their specifications, \ie docstring.
However, \toolname{} successfully queries ChatGPT to produce the correct code for the first example with code search, the second example with program repair, and the third example with both code search and program repair. 

\begin{figure}
    \centering
    \includegraphics[width=0.7\linewidth]{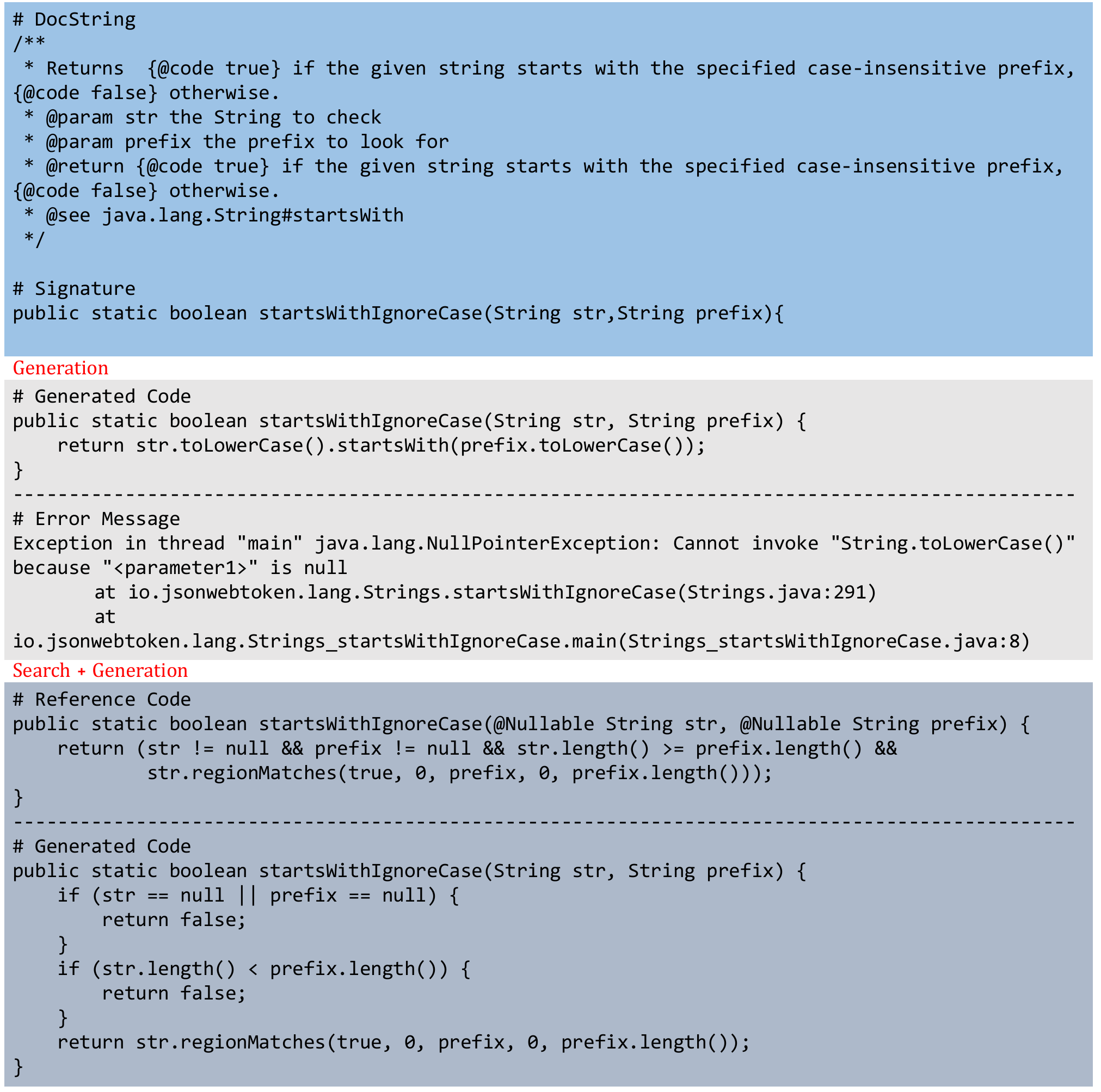}
    \caption{Example-1 generated by ChatGPT}
    \label{fig:case1}
\end{figure}

\begin{figure}
    \centering
    \includegraphics[width=0.7\linewidth]{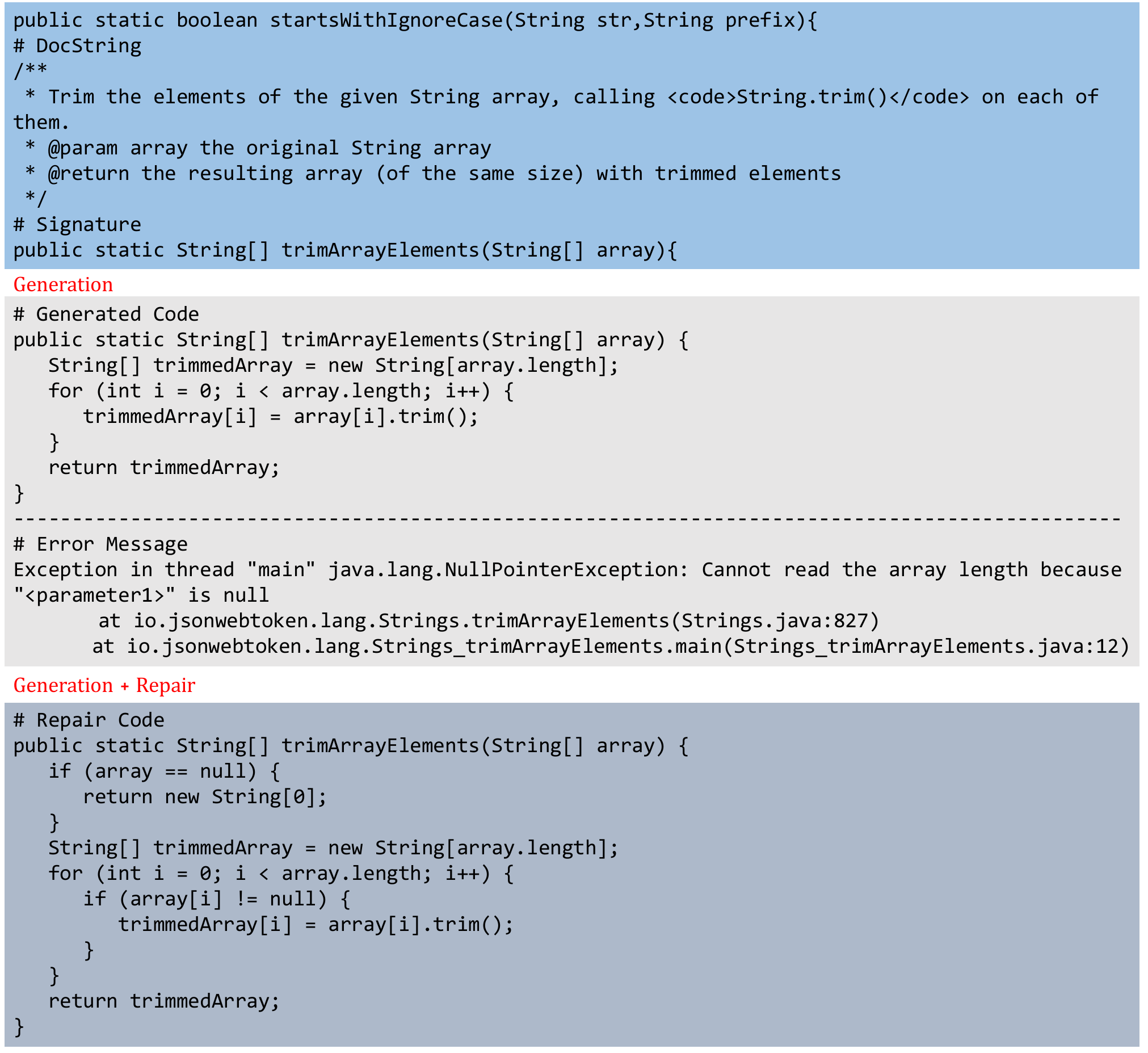}
    \caption{Example-2 generated by ChatGPT}
    \label{fig:case2}
\end{figure}

\begin{figure}
    \centering
    \includegraphics[width=0.7\linewidth]{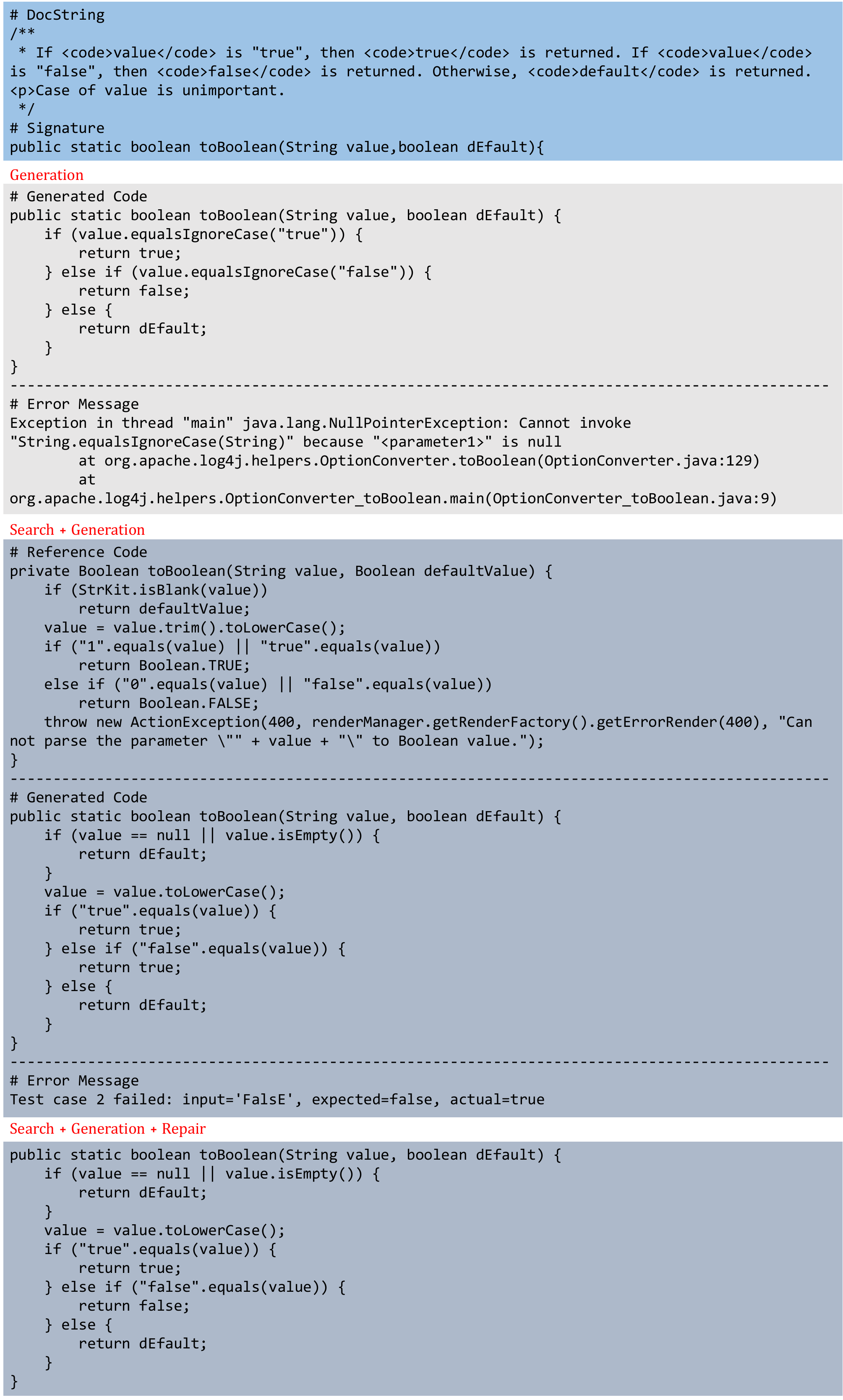}
    \caption{Example-3 generated by ChatGPT}
    \label{fig:case3}
\end{figure}

\textbf{Case 1}.
Fig.~\ref{fig:case1} illustrates an example that is not correctly generated by ChatGPT, but correct code can be produced by providing relevant information from code search.
This example attempts to identify whether a given string starts with a specified case-insensitive prefix, as shown in the blue part of Fig.~\ref{fig:case1}.
ChatGPT first attempts to generate the solution directly, which ignores some boundary conditions and leads to a `NullPointerException` if either ``str'' or ``prefix'' is null.
\toolname{} then retrieves a similar solution that provides additional context and correct implementation details, with which ChatGPT properly handles ``null'' values and checks the length of the strings before performing the comparison, thus avoiding the ``NullPointerException''.
Overall, with the guidance of retrieved code, LLMs produce higher-quality code that adheres to best practices and avoids common pitfalls. 

\textbf{Case 2}.
Fig.~\ref{fig:case2} illustrates an example that fails in the generation-only scenario but succeeds after program repair.
This example requires completing the logic to call ``String.trim()'' on each element in a given string array.
Similar to the generate-with-retrieval case, CHatGPT overlooks boundary conditions and fails the null-input test case. 
When provided with the dynamic error message, ChatGPT fixes the bug related to missing boundary checks and passes all test cases.

\textbf{Case 3}.
Fig.~\ref{fig:case3} illustrates an example only generated corrected with code search and program repair.
This example involves implementing logic that returns true when the given value is ``true'', returns false when the given value is ``false'', and otherwise returns a specified default value. 
ChatGPT first attempts to generate the solution while ignoring the possibility of a null input, thus causing a NullPointerException. 
After retrieving the reference code, ChatGPT identifies the error and corrects the faulty logic. 
However, it encounters a new issue where it fails to return the correct value in the ``false'' condition for the same reason. 
Finally equipped with program repair, ChatGPT generates the correct code and finally passes the test.

\section{Discussion and Future Work}
\label{sec:dis}
The primary technical innovation in this work is the introduction of a unified automatic programming paradigm that leverages advanced LLMs to integrate three long-explored research areas, \ie code search code generation and program repair.
The preliminary experiments highlight the potential of \toolname{} in competitive programming problems and real-world programming scenarios.
Particularly, we demonstrate that (1) code search can help code generators produce more accurate code; (2) program repair serves as an effective post-processing step, even after retrieval-augmented code generation; (3) a unified programming pipeline, incorporating the above three phases together, is highly effective in generating code, especially when equipped with LLMs.

As a unified programming pipeline, we believe \toolname{} has significant potential for the SE community, and can be extended in the following aspects.

\textbf{Deployment Scenarios}.
It is promising to adapt \toolname{} to more programming scenarios during deployment.
First, there are some domain-specific areas that require developers to possess both programming skills and domain expert knowledge, such as hardware code~\cite{ahmad2024hardware}.
\toolname{} can fully automate such programming process by retrieving similar code within the domain and iteratively refining it, thereby reducing the programming barrier for developers.
Second, \toolname{} takes natural language descriptions as inputs currently, but it can be implemented with other query formats, such as test cases.
Considering that \toolname{} treats queries as tokens or embeddings without taking any specific code features into account, \toolname{} can be applied to the other input formats in a drop-in fashion.
For example, the potential of \toolname{} in the well-known test-driven development~\cite{rafique2013effects} is worth investigating, \ie, retrieving similar solutions based on test cases.
Third, \toolname{} is generic to other code-related tasks, such as test generation, code translation and program repair.

\textbf{Technical Designs}.
The effectiveness of \toolname{} can be further optimized by improving the quality of retrieved code and refining the prompt engineering for program repair.
First, \toolname{} employs a straightforward retriever that can be either token-based or embedding-based, depending on whether it focuses on syntax or semantics.
Given the rapid advancements in code search, optimizing these retrieval strategies for greater efficiency is essential.
Second, \toolname{} directly appends error messages to the generated code as prompts and then queries LLMs for repair. 
In the future, more advanced prompt engineering techniques, such as chain-of-thought, can be utilized.

\textbf{Evaluation Experiments}.
In this study, preliminary experiments demonstrate the effectiveness of \toolname{}.
We only conduct case studies in Section~\ref{sec:rq2} due to the limitation of real-world programming datasets and retrieval corpora.
Large-scale evaluations with quantitative analysis are necessary in the future.
Besides, the opinions and experiences of developers are crucial for assessing the utility of such programming tools.
We plan to conduct user studies to evaluate \toolname{} in real-world programming scenarios.
Furthermore, the experiments can be extended in the future with more studied LLMs, benchmarks, and programming languages.

\section{Conclusion}
\label{sec:conclusion}
In this paper, we propose a novel automatic programming framework, \toolname{}, which leverages advanced large language models (LLMs) to integrate three well-established areas: code search, code generation, and program repair. 
Preliminary experiments indicate the potential of our framework to enhance the problem-solving capabilities of existing LLMs in programming tasks. 
Besides, our framework demonstrates the preliminary benefits of combining LLMs with traditional software engineering (SE) areas. 
In the future, more advanced technologies, such as intelligent agents, can be employed to further integrate various SE techniques within the programming framework more effectively.

\bibliographystyle{ACM-Reference-Format}
\bibliography{reference}


\begin{thebibliography}{51}


\ifx \showCODEN    \undefined \def \showCODEN     #1{\unskip}     \fi
\ifx \showDOI      \undefined \def \showDOI       #1{#1}\fi
\ifx \showISBNx    \undefined \def \showISBNx     #1{\unskip}     \fi
\ifx \showISBNxiii \undefined \def \showISBNxiii  #1{\unskip}     \fi
\ifx \showISSN     \undefined \def \showISSN      #1{\unskip}     \fi
\ifx \showLCCN     \undefined \def \showLCCN      #1{\unskip}     \fi
\ifx \shownote     \undefined \def \shownote      #1{#1}          \fi
\ifx \showarticletitle \undefined \def \showarticletitle #1{#1}   \fi
\ifx \showURL      \undefined \def \showURL       {\relax}        \fi
\providecommand\bibfield[2]{#2}
\providecommand\bibinfo[2]{#2}
\providecommand\natexlab[1]{#1}
\providecommand\showeprint[2][]{arXiv:#2}

\bibitem[Ahmad et~al\mbox{.}(2024)]%
        {ahmad2024hardware}
\bibfield{author}{\bibinfo{person}{Baleegh Ahmad}, \bibinfo{person}{Shailja Thakur}, \bibinfo{person}{Benjamin Tan}, \bibinfo{person}{Ramesh Karri}, {and} \bibinfo{person}{Hammond Pearce}.} \bibinfo{year}{2024}\natexlab{}.
\newblock \showarticletitle{On Hardware Security Bug Code Fixes by Prompting Large Language Models}.
\newblock \bibinfo{journal}{\emph{{IEEE} Trans. Inf. Forensics Secur.}}  \bibinfo{volume}{19} (\bibinfo{year}{2024}), \bibinfo{pages}{4043--4057}.
\newblock
\urldef\tempurl%
\url{https://doi.org/10.1109/TIFS.2024.3374558}
\showDOI{\tempurl}


\bibitem[Austin et~al\mbox{.}(2021)]%
        {austin2021program}
\bibfield{author}{\bibinfo{person}{Jacob Austin}, \bibinfo{person}{Augustus Odena}, \bibinfo{person}{Maxwell~I. Nye}, \bibinfo{person}{Maarten Bosma}, \bibinfo{person}{Henryk Michalewski}, \bibinfo{person}{David Dohan}, \bibinfo{person}{Ellen Jiang}, \bibinfo{person}{Carrie~J. Cai}, \bibinfo{person}{Michael Terry}, \bibinfo{person}{Quoc~V. Le}, {and} \bibinfo{person}{Charles Sutton}.} \bibinfo{year}{2021}\natexlab{}.
\newblock \showarticletitle{Program Synthesis with Large Language Models}.
\newblock \bibinfo{journal}{\emph{CoRR}}  \bibinfo{volume}{abs/2108.07732} (\bibinfo{year}{2021}).
\newblock
\showeprint[arXiv]{2108.07732}
\urldef\tempurl%
\url{https://arxiv.org/abs/2108.07732}
\showURL{%
\tempurl}


\bibitem[Barr et~al\mbox{.}(2014)]%
        {barr2022plastic}
\bibfield{author}{\bibinfo{person}{Earl~T. Barr}, \bibinfo{person}{Yuriy Brun}, \bibinfo{person}{Premkumar~T. Devanbu}, \bibinfo{person}{Mark Harman}, {and} \bibinfo{person}{Federica Sarro}.} \bibinfo{year}{2014}\natexlab{}.
\newblock \showarticletitle{The plastic surgery hypothesis}. In \bibinfo{booktitle}{\emph{Proceedings of the 22nd {ACM} {SIGSOFT} International Symposium on Foundations of Software Engineering, (FSE-22), Hong Kong, China, November 16 - 22, 2014}}, \bibfield{editor}{\bibinfo{person}{Shing{-}Chi Cheung}, \bibinfo{person}{Alessandro Orso}, {and} \bibinfo{person}{Margaret{-}Anne~D. Storey}} (Eds.). \bibinfo{publisher}{{ACM}}, \bibinfo{pages}{306--317}.
\newblock
\urldef\tempurl%
\url{https://doi.org/10.1145/2635868.2635898}
\showDOI{\tempurl}


\bibitem[Chen et~al\mbox{.}(2023)]%
        {chen2023teaching}
\bibfield{author}{\bibinfo{person}{Xinyun Chen}, \bibinfo{person}{Maxwell Lin}, \bibinfo{person}{Nathanael Sch{\"{a}}rli}, {and} \bibinfo{person}{Denny Zhou}.} \bibinfo{year}{2023}\natexlab{}.
\newblock \showarticletitle{Teaching Large Language Models to Self-Debug}.
\newblock \bibinfo{journal}{\emph{CoRR}}  \bibinfo{volume}{abs/2304.05128} (\bibinfo{year}{2023}).
\newblock
\urldef\tempurl%
\url{https://doi.org/10.48550/ARXIV.2304.05128}
\showDOI{\tempurl}
\showeprint[arXiv]{2304.05128}


\bibitem[Du et~al\mbox{.}(2024)]%
        {du2023classeval}
\bibfield{author}{\bibinfo{person}{Xueying Du}, \bibinfo{person}{Mingwei Liu}, \bibinfo{person}{Kaixin Wang}, \bibinfo{person}{Hanlin Wang}, \bibinfo{person}{Junwei Liu}, \bibinfo{person}{Yixuan Chen}, \bibinfo{person}{Jiayi Feng}, \bibinfo{person}{Chaofeng Sha}, \bibinfo{person}{Xin Peng}, {and} \bibinfo{person}{Yiling Lou}.} \bibinfo{year}{2024}\natexlab{}.
\newblock \showarticletitle{Evaluating Large Language Models in Class-Level Code Generation}. In \bibinfo{booktitle}{\emph{Proceedings of the 46th {IEEE/ACM} International Conference on Software Engineering, {ICSE} 2024, Lisbon, Portugal, April 14-20, 2024}}. \bibinfo{publisher}{{ACM}}, \bibinfo{pages}{81:1--81:13}.
\newblock
\urldef\tempurl%
\url{https://doi.org/10.1145/3597503.3639219}
\showDOI{\tempurl}


\bibitem[Fan et~al\mbox{.}(2023)]%
        {fan2023automated}
\bibfield{author}{\bibinfo{person}{Zhiyu Fan}, \bibinfo{person}{Xiang Gao}, \bibinfo{person}{Martin Mirchev}, \bibinfo{person}{Abhik Roychoudhury}, {and} \bibinfo{person}{Shin~Hwei Tan}.} \bibinfo{year}{2023}\natexlab{}.
\newblock \showarticletitle{Automated Repair of Programs from Large Language Models}. In \bibinfo{booktitle}{\emph{45th {IEEE/ACM} International Conference on Software Engineering, {ICSE} 2023, Melbourne, Australia, May 14-20, 2023}}. \bibinfo{publisher}{{IEEE}}, \bibinfo{pages}{1469--1481}.
\newblock
\urldef\tempurl%
\url{https://doi.org/10.1109/ICSE48619.2023.00128}
\showDOI{\tempurl}


\bibitem[Feng et~al\mbox{.}(2020)]%
        {2020fengcodebert}
\bibfield{author}{\bibinfo{person}{Zhangyin Feng}, \bibinfo{person}{Daya Guo}, \bibinfo{person}{Duyu Tang}, \bibinfo{person}{Nan Duan}, \bibinfo{person}{Xiaocheng Feng}, \bibinfo{person}{Ming Gong}, \bibinfo{person}{Linjun Shou}, \bibinfo{person}{Bing Qin}, \bibinfo{person}{Ting Liu}, \bibinfo{person}{Daxin Jiang}, {and} \bibinfo{person}{Ming Zhou}.} \bibinfo{year}{2020}\natexlab{}.
\newblock \showarticletitle{CodeBERT: {A} Pre-Trained Model for Programming and Natural Languages}. In \bibinfo{booktitle}{\emph{Findings of the Association for Computational Linguistics: {EMNLP} 2020, Online Event, 16-20 November 2020}} \emph{(\bibinfo{series}{Findings of {ACL}}, Vol.~\bibinfo{volume}{{EMNLP} 2020})}, \bibfield{editor}{\bibinfo{person}{Trevor Cohn}, \bibinfo{person}{Yulan He}, {and} \bibinfo{person}{Yang Liu}} (Eds.). \bibinfo{publisher}{Association for Computational Linguistics}, \bibinfo{pages}{1536--1547}.
\newblock
\urldef\tempurl%
\url{https://doi.org/10.18653/V1/2020.FINDINGS-EMNLP.139}
\showDOI{\tempurl}


\bibitem[Foundation(2024)]%
        {apache_lucene}
\bibfield{author}{\bibinfo{person}{Apache~Software Foundation}.} \bibinfo{year}{2024}\natexlab{}.
\newblock \bibinfo{title}{Apache Lucene}.
\newblock \bibinfo{howpublished}{\url{https://lucene.apache.org}}.
\newblock
\newblock
\shownote{Accessed: 2024-07-29}.


\bibitem[Gazzola et~al\mbox{.}(2019)]%
        {gazzola2019automatic}
\bibfield{author}{\bibinfo{person}{Luca Gazzola}, \bibinfo{person}{Daniela Micucci}, {and} \bibinfo{person}{Leonardo Mariani}.} \bibinfo{year}{2019}\natexlab{}.
\newblock \showarticletitle{Automatic Software Repair: {A} Survey}.
\newblock \bibinfo{journal}{\emph{{IEEE} Trans. Software Eng.}} \bibinfo{volume}{45}, \bibinfo{number}{1} (\bibinfo{year}{2019}), \bibinfo{pages}{34--67}.
\newblock
\urldef\tempurl%
\url{https://doi.org/10.1109/TSE.2017.2755013}
\showDOI{\tempurl}


\bibitem[Grazia and Pradel(2023)]%
        {di2023code}
\bibfield{author}{\bibinfo{person}{Luca~Di Grazia} {and} \bibinfo{person}{Michael Pradel}.} \bibinfo{year}{2023}\natexlab{}.
\newblock \showarticletitle{Code Search: {A} Survey of Techniques for Finding Code}.
\newblock \bibinfo{journal}{\emph{{ACM} Comput. Surv.}} \bibinfo{volume}{55}, \bibinfo{number}{11} (\bibinfo{year}{2023}), \bibinfo{pages}{220:1--220:31}.
\newblock
\urldef\tempurl%
\url{https://doi.org/10.1145/3565971}
\showDOI{\tempurl}


\bibitem[Jiang et~al\mbox{.}(2024)]%
        {jiang2024survey}
\bibfield{author}{\bibinfo{person}{Juyong Jiang}, \bibinfo{person}{Fan Wang}, \bibinfo{person}{Jiasi Shen}, \bibinfo{person}{Sungju Kim}, {and} \bibinfo{person}{Sunghun Kim}.} \bibinfo{year}{2024}\natexlab{}.
\newblock \showarticletitle{A Survey on Large Language Models for Code Generation}.
\newblock \bibinfo{journal}{\emph{CoRR}}  \bibinfo{volume}{abs/2406.00515} (\bibinfo{year}{2024}).
\newblock
\urldef\tempurl%
\url{https://doi.org/10.48550/ARXIV.2406.00515}
\showDOI{\tempurl}
\showeprint[arXiv]{2406.00515}


\bibitem[Jiang et~al\mbox{.}(2023)]%
        {jiang2023impact}
\bibfield{author}{\bibinfo{person}{Nan Jiang}, \bibinfo{person}{Kevin Liu}, \bibinfo{person}{Thibaud Lutellier}, {and} \bibinfo{person}{Lin Tan}.} \bibinfo{year}{2023}\natexlab{}.
\newblock \showarticletitle{Impact of Code Language Models on Automated Program Repair}. In \bibinfo{booktitle}{\emph{45th {IEEE/ACM} International Conference on Software Engineering, {ICSE} 2023, Melbourne, Australia, May 14-20, 2023}}. \bibinfo{publisher}{{IEEE}}, \bibinfo{pages}{1430--1442}.
\newblock
\urldef\tempurl%
\url{https://doi.org/10.1109/ICSE48619.2023.00125}
\showDOI{\tempurl}


\bibitem[Just et~al\mbox{.}(2014)]%
        {just2014defects4j}
\bibfield{author}{\bibinfo{person}{Ren{\'{e}} Just}, \bibinfo{person}{Darioush Jalali}, {and} \bibinfo{person}{Michael~D. Ernst}.} \bibinfo{year}{2014}\natexlab{}.
\newblock \showarticletitle{Defects4J: a database of existing faults to enable controlled testing studies for Java programs}. In \bibinfo{booktitle}{\emph{International Symposium on Software Testing and Analysis, {ISSTA} '14, San Jose, CA, {USA} - July 21 - 26, 2014}}, \bibfield{editor}{\bibinfo{person}{Corina~S. Pasareanu} {and} \bibinfo{person}{Darko Marinov}} (Eds.). \bibinfo{publisher}{{ACM}}, \bibinfo{pages}{437--440}.
\newblock
\urldef\tempurl%
\url{https://doi.org/10.1145/2610384.2628055}
\showDOI{\tempurl}


\bibitem[Kim et~al\mbox{.}(2024)]%
        {kim2024big}
\bibfield{author}{\bibinfo{person}{Kisub Kim}, \bibinfo{person}{Sankalp Ghatpande}, \bibinfo{person}{Dongsun Kim}, \bibinfo{person}{Xin Zhou}, \bibinfo{person}{Kui Liu}, \bibinfo{person}{Tegawend{\'{e}}~F. Bissyand{\'{e}}}, \bibinfo{person}{Jacques Klein}, {and} \bibinfo{person}{Yves~Le Traon}.} \bibinfo{year}{2024}\natexlab{}.
\newblock \showarticletitle{Big Code Search: {A} Bibliography}.
\newblock \bibinfo{journal}{\emph{{ACM} Comput. Surv.}} \bibinfo{volume}{56}, \bibinfo{number}{1} (\bibinfo{year}{2024}), \bibinfo{pages}{25:1--25:49}.
\newblock
\urldef\tempurl%
\url{https://doi.org/10.1145/3604905}
\showDOI{\tempurl}


\bibitem[{Le Goues} et~al\mbox{.}(2019)]%
        {goues2019automated}
\bibfield{author}{\bibinfo{person}{Claire {Le Goues}}, \bibinfo{person}{Michael Pradel}, {and} \bibinfo{person}{Abhik Roychoudhury}.} \bibinfo{year}{2019}\natexlab{}.
\newblock \showarticletitle{Automated program repair}.
\newblock \bibinfo{journal}{\emph{Commun. {ACM}}} \bibinfo{volume}{62}, \bibinfo{number}{12} (\bibinfo{year}{2019}), \bibinfo{pages}{56--65}.
\newblock
\urldef\tempurl%
\url{https://doi.org/10.1145/3318162}
\showDOI{\tempurl}


\bibitem[Li et~al\mbox{.}(2023)]%
        {li2023skcoder}
\bibfield{author}{\bibinfo{person}{Jia Li}, \bibinfo{person}{Yongmin Li}, \bibinfo{person}{Ge Li}, \bibinfo{person}{Zhi Jin}, \bibinfo{person}{Yiyang Hao}, {and} \bibinfo{person}{Xing Hu}.} \bibinfo{year}{2023}\natexlab{}.
\newblock \showarticletitle{SkCoder: {A} Sketch-based Approach for Automatic Code Generation}. In \bibinfo{booktitle}{\emph{45th {IEEE/ACM} International Conference on Software Engineering, {ICSE} 2023, Melbourne, Australia, May 14-20, 2023}}. \bibinfo{publisher}{{IEEE}}, \bibinfo{pages}{2124--2135}.
\newblock
\urldef\tempurl%
\url{https://doi.org/10.1109/ICSE48619.2023.00179}
\showDOI{\tempurl}


\bibitem[Li et~al\mbox{.}(2024)]%
        {li2024acecoder}
\bibfield{author}{\bibinfo{person}{Jia Li}, \bibinfo{person}{Yunfei Zhao}, \bibinfo{person}{Yongmin Li}, \bibinfo{person}{Ge Li}, {and} \bibinfo{person}{Zhi Jin}.} \bibinfo{year}{2024}\natexlab{}.
\newblock \showarticletitle{AceCoder: An Effective Prompting Technique Specialized in Code Generation}.
\newblock \bibinfo{journal}{\emph{ACM Trans. Softw. Eng. Methodol.}} (\bibinfo{date}{jul} \bibinfo{year}{2024}).
\newblock
\showISSN{1049-331X}
\urldef\tempurl%
\url{https://doi.org/10.1145/3675395}
\showDOI{\tempurl}
\newblock
\shownote{Just Accepted}.


\bibitem[Liu et~al\mbox{.}(2022)]%
        {liu2022opportunities}
\bibfield{author}{\bibinfo{person}{Chao Liu}, \bibinfo{person}{Xin Xia}, \bibinfo{person}{David Lo}, \bibinfo{person}{Cuiyun Gao}, \bibinfo{person}{Xiaohu Yang}, {and} \bibinfo{person}{John~C. Grundy}.} \bibinfo{year}{2022}\natexlab{}.
\newblock \showarticletitle{Opportunities and Challenges in Code Search Tools}.
\newblock \bibinfo{journal}{\emph{{ACM} Comput. Surv.}} \bibinfo{volume}{54}, \bibinfo{number}{9} (\bibinfo{year}{2022}), \bibinfo{pages}{196:1--196:40}.
\newblock
\urldef\tempurl%
\url{https://doi.org/10.1145/3480027}
\showDOI{\tempurl}


\bibitem[Liu et~al\mbox{.}(2023)]%
        {liu2023graphsearchnet}
\bibfield{author}{\bibinfo{person}{Shangqing Liu}, \bibinfo{person}{Xiaofei Xie}, \bibinfo{person}{Jing~Kai Siow}, \bibinfo{person}{Lei Ma}, \bibinfo{person}{Guozhu Meng}, {and} \bibinfo{person}{Yang Liu}.} \bibinfo{year}{2023}\natexlab{}.
\newblock \showarticletitle{GraphSearchNet: Enhancing GNNs via Capturing Global Dependencies for Semantic Code Search}.
\newblock \bibinfo{journal}{\emph{{IEEE} Trans. Software Eng.}} \bibinfo{volume}{49}, \bibinfo{number}{4} (\bibinfo{year}{2023}), \bibinfo{pages}{2839--2855}.
\newblock
\urldef\tempurl%
\url{https://doi.org/10.1109/TSE.2022.3233901}
\showDOI{\tempurl}


\bibitem[Liu et~al\mbox{.}(2024)]%
        {liu2024refining}
\bibfield{author}{\bibinfo{person}{Yue Liu}, \bibinfo{person}{Thanh Le{-}Cong}, \bibinfo{person}{Ratnadira Widyasari}, \bibinfo{person}{Chakkrit Tantithamthavorn}, \bibinfo{person}{Li Li}, \bibinfo{person}{Xuan{-}Bach~Dinh Le}, {and} \bibinfo{person}{David Lo}.} \bibinfo{year}{2024}\natexlab{}.
\newblock \showarticletitle{Refining ChatGPT-Generated Code: Characterizing and Mitigating Code Quality Issues}.
\newblock \bibinfo{journal}{\emph{{ACM} Trans. Softw. Eng. Methodol.}} \bibinfo{volume}{33}, \bibinfo{number}{5} (\bibinfo{year}{2024}), \bibinfo{pages}{116:1--116:26}.
\newblock
\urldef\tempurl%
\url{https://doi.org/10.1145/3643674}
\showDOI{\tempurl}


\bibitem[Lutellier et~al\mbox{.}(2020)]%
        {lutellier2020coconut}
\bibfield{author}{\bibinfo{person}{Thibaud Lutellier}, \bibinfo{person}{Hung~Viet Pham}, \bibinfo{person}{Lawrence Pang}, \bibinfo{person}{Yitong Li}, \bibinfo{person}{Moshi Wei}, {and} \bibinfo{person}{Lin Tan}.} \bibinfo{year}{2020}\natexlab{}.
\newblock \showarticletitle{CoCoNuT: combining context-aware neural translation models using ensemble for program repair}. In \bibinfo{booktitle}{\emph{{ISSTA} '20: 29th {ACM} {SIGSOFT} International Symposium on Software Testing and Analysis, Virtual Event, USA, July 18-22, 2020}}, \bibfield{editor}{\bibinfo{person}{Sarfraz Khurshid} {and} \bibinfo{person}{Corina~S. Pasareanu}} (Eds.). \bibinfo{publisher}{{ACM}}, \bibinfo{pages}{101--114}.
\newblock
\urldef\tempurl%
\url{https://doi.org/10.1145/3395363.3397369}
\showDOI{\tempurl}


\bibitem[Mastropaolo et~al\mbox{.}(2023)]%
        {mastropaolo2023robustness}
\bibfield{author}{\bibinfo{person}{Antonio Mastropaolo}, \bibinfo{person}{Luca Pascarella}, \bibinfo{person}{Emanuela Guglielmi}, \bibinfo{person}{Matteo Ciniselli}, \bibinfo{person}{Simone Scalabrino}, \bibinfo{person}{Rocco Oliveto}, {and} \bibinfo{person}{Gabriele Bavota}.} \bibinfo{year}{2023}\natexlab{}.
\newblock \showarticletitle{On the Robustness of Code Generation Techniques: An Empirical Study on GitHub Copilot}. In \bibinfo{booktitle}{\emph{45th {IEEE/ACM} International Conference on Software Engineering, {ICSE} 2023, Melbourne, Australia, May 14-20, 2023}}. \bibinfo{publisher}{{IEEE}}, \bibinfo{pages}{2149--2160}.
\newblock
\urldef\tempurl%
\url{https://doi.org/10.1109/ICSE48619.2023.00181}
\showDOI{\tempurl}


\bibitem[Mei and Zhang(2018)]%
        {mei2018can}
\bibfield{author}{\bibinfo{person}{Hong Mei} {and} \bibinfo{person}{Lu Zhang}.} \bibinfo{year}{2018}\natexlab{}.
\newblock \showarticletitle{Can big data bring a breakthrough for software automation?}
\newblock \bibinfo{journal}{\emph{Sci. China Inf. Sci.}} \bibinfo{volume}{61}, \bibinfo{number}{5} (\bibinfo{year}{2018}), \bibinfo{pages}{056101:1--056101:3}.
\newblock
\urldef\tempurl%
\url{https://doi.org/10.1007/S11432-017-9355-3}
\showDOI{\tempurl}


\bibitem[Monperrus(2018)]%
        {monperrus2019automatic}
\bibfield{author}{\bibinfo{person}{Martin Monperrus}.} \bibinfo{year}{2018}\natexlab{}.
\newblock \showarticletitle{Automatic Software Repair: {A} Bibliography}.
\newblock \bibinfo{journal}{\emph{{ACM} Comput. Surv.}} \bibinfo{volume}{51}, \bibinfo{number}{1} (\bibinfo{year}{2018}), \bibinfo{pages}{17:1--17:24}.
\newblock
\urldef\tempurl%
\url{https://doi.org/10.1145/3105906}
\showDOI{\tempurl}


\bibitem[Olausson et~al\mbox{.}(2023)]%
        {olausson2023self}
\bibfield{author}{\bibinfo{person}{Theo~X Olausson}, \bibinfo{person}{Jeevana~Priya Inala}, \bibinfo{person}{Chenglong Wang}, \bibinfo{person}{Jianfeng Gao}, {and} \bibinfo{person}{Armando Solar-Lezama}.} \bibinfo{year}{2023}\natexlab{}.
\newblock \showarticletitle{Is Self-Repair a Silver Bullet for Code Generation?}. In \bibinfo{booktitle}{\emph{The Twelfth International Conference on Learning Representations}}.
\newblock


\bibitem[OpenAI(2023)]%
        {gpt4}
\bibfield{author}{\bibinfo{person}{OpenAI}.} \bibinfo{year}{2023}\natexlab{}.
\newblock \showarticletitle{{GPT-4} Technical Report}.
\newblock \bibinfo{journal}{\emph{CoRR}}  \bibinfo{volume}{abs/2303.08774} (\bibinfo{year}{2023}).
\newblock
\urldef\tempurl%
\url{https://doi.org/10.48550/ARXIV.2303.08774}
\showDOI{\tempurl}
\showeprint[arXiv]{2303.08774}


\bibitem[OpenAI(2024)]%
        {chatgpt}
\bibfield{author}{\bibinfo{person}{OpenAI}.} \bibinfo{year}{2024}\natexlab{}.
\newblock \bibinfo{title}{{{ChatGPT}}}.
\newblock \bibinfo{howpublished}{URL: \url{https://openai.com/blog/ChatGPT}}.
\newblock


\bibitem[Rafique and Misic(2013)]%
        {rafique2013effects}
\bibfield{author}{\bibinfo{person}{Yahya Rafique} {and} \bibinfo{person}{Vojislav~B. Misic}.} \bibinfo{year}{2013}\natexlab{}.
\newblock \showarticletitle{The Effects of Test-Driven Development on External Quality and Productivity: {A} Meta-Analysis}.
\newblock \bibinfo{journal}{\emph{{IEEE} Trans. Software Eng.}} \bibinfo{volume}{39}, \bibinfo{number}{6} (\bibinfo{year}{2013}), \bibinfo{pages}{835--856}.
\newblock
\urldef\tempurl%
\url{https://doi.org/10.1109/TSE.2012.28}
\showDOI{\tempurl}


\bibitem[Roychoudhury and Xiong(2019)]%
        {roychoudhury2019automated}
\bibfield{author}{\bibinfo{person}{Abhik Roychoudhury} {and} \bibinfo{person}{Yingfei Xiong}.} \bibinfo{year}{2019}\natexlab{}.
\newblock \showarticletitle{Automated program repair: a step towards software automation}.
\newblock \bibinfo{journal}{\emph{Sci. China Inf. Sci.}} \bibinfo{volume}{62}, \bibinfo{number}{10} (\bibinfo{year}{2019}), \bibinfo{pages}{200103:1--200103:3}.
\newblock
\urldef\tempurl%
\url{https://doi.org/10.1007/S11432-019-9947-6}
\showDOI{\tempurl}


\bibitem[Rozi{\`{e}}re et~al\mbox{.}(2023)]%
        {codellama}
\bibfield{author}{\bibinfo{person}{Baptiste Rozi{\`{e}}re}, \bibinfo{person}{Jonas Gehring}, \bibinfo{person}{Fabian Gloeckle}, \bibinfo{person}{Sten Sootla}, \bibinfo{person}{Itai Gat}, \bibinfo{person}{Xiaoqing~Ellen Tan}, \bibinfo{person}{Yossi Adi}, \bibinfo{person}{Jingyu Liu}, \bibinfo{person}{Tal Remez}, \bibinfo{person}{J{\'{e}}r{\'{e}}my Rapin}, \bibinfo{person}{Artyom Kozhevnikov}, \bibinfo{person}{Ivan Evtimov}, \bibinfo{person}{Joanna Bitton}, \bibinfo{person}{Manish Bhatt}, \bibinfo{person}{Cristian Canton{-}Ferrer}, \bibinfo{person}{Aaron Grattafiori}, \bibinfo{person}{Wenhan Xiong}, \bibinfo{person}{Alexandre D{\'{e}}fossez}, \bibinfo{person}{Jade Copet}, \bibinfo{person}{Faisal Azhar}, \bibinfo{person}{Hugo Touvron}, \bibinfo{person}{Louis Martin}, \bibinfo{person}{Nicolas Usunier}, \bibinfo{person}{Thomas Scialom}, {and} \bibinfo{person}{Gabriel Synnaeve}.} \bibinfo{year}{2023}\natexlab{}.
\newblock \showarticletitle{Code Llama: Open Foundation Models for Code}.
\newblock \bibinfo{journal}{\emph{CoRR}}  \bibinfo{volume}{abs/2308.12950} (\bibinfo{year}{2023}).
\newblock
\urldef\tempurl%
\url{https://doi.org/10.48550/ARXIV.2308.12950}
\showDOI{\tempurl}
\showeprint[arXiv]{2308.12950}


\bibitem[Sun et~al\mbox{.}(2024)]%
        {sun2024survey}
\bibfield{author}{\bibinfo{person}{Weisong Sun}, \bibinfo{person}{Chunrong Fang}, \bibinfo{person}{Yifei Ge}, \bibinfo{person}{Yuling Hu}, \bibinfo{person}{Yuchen Chen}, \bibinfo{person}{Quanjun Zhang}, \bibinfo{person}{Xiuting Ge}, \bibinfo{person}{Yang Liu}, {and} \bibinfo{person}{Zhenyu Chen}.} \bibinfo{year}{2024}\natexlab{}.
\newblock \showarticletitle{A Survey of Source Code Search: A 3-Dimensional Perspective}.
\newblock \bibinfo{journal}{\emph{ACM Trans. Softw. Eng. Methodol.}} (\bibinfo{date}{apr} \bibinfo{year}{2024}).
\newblock
\showISSN{1049-331X}
\urldef\tempurl%
\url{https://doi.org/10.1145/3656341}
\showDOI{\tempurl}
\newblock
\shownote{Just Accepted}.


\bibitem[Tang et~al\mbox{.}(2024)]%
        {tang2024codeagent}
\bibfield{author}{\bibinfo{person}{Daniel Tang}, \bibinfo{person}{Zhenghan Chen}, \bibinfo{person}{Kisub Kim}, \bibinfo{person}{Yewei Song}, \bibinfo{person}{Haoye Tian}, \bibinfo{person}{Saad Ezzini}, \bibinfo{person}{Yongfeng Huang}, \bibinfo{person}{Jacques Klein}, {and} \bibinfo{person}{Tegawend{\'{e}}~F. Bissyand{\'{e}}}.} \bibinfo{year}{2024}\natexlab{}.
\newblock \showarticletitle{CodeAgent: Collaborative Agents for Software Engineering}.
\newblock \bibinfo{journal}{\emph{CoRR}}  \bibinfo{volume}{abs/2402.02172} (\bibinfo{year}{2024}).
\newblock
\urldef\tempurl%
\url{https://doi.org/10.48550/ARXIV.2402.02172}
\showDOI{\tempurl}
\showeprint[arXiv]{2402.02172}


\bibitem[Tufano et~al\mbox{.}(2019)]%
        {tufano2019empirical}
\bibfield{author}{\bibinfo{person}{Michele Tufano}, \bibinfo{person}{Cody Watson}, \bibinfo{person}{Gabriele Bavota}, \bibinfo{person}{Massimiliano~Di Penta}, \bibinfo{person}{Martin White}, {and} \bibinfo{person}{Denys Poshyvanyk}.} \bibinfo{year}{2019}\natexlab{}.
\newblock \showarticletitle{An Empirical Study on Learning Bug-Fixing Patches in the Wild via Neural Machine Translation}.
\newblock \bibinfo{journal}{\emph{{ACM} Trans. Softw. Eng. Methodol.}} \bibinfo{volume}{28}, \bibinfo{number}{4} (\bibinfo{year}{2019}), \bibinfo{pages}{19:1--19:29}.
\newblock
\urldef\tempurl%
\url{https://doi.org/10.1145/3340544}
\showDOI{\tempurl}


\bibitem[Wang et~al\mbox{.}(2023a)]%
        {wang2023scientific}
\bibfield{author}{\bibinfo{person}{Hanchen Wang}, \bibinfo{person}{Tianfan Fu}, \bibinfo{person}{Yuanqi Du}, \bibinfo{person}{Wenhao Gao}, \bibinfo{person}{Kexin Huang}, \bibinfo{person}{Ziming Liu}, \bibinfo{person}{Payal Chandak}, \bibinfo{person}{Shengchao Liu}, \bibinfo{person}{Peter~Van Katwyk}, \bibinfo{person}{Andreea Deac}, \bibinfo{person}{Anima Anandkumar}, \bibinfo{person}{Karianne Bergen}, \bibinfo{person}{Carla~P. Gomes}, \bibinfo{person}{Shirley Ho}, \bibinfo{person}{Pushmeet Kohli}, \bibinfo{person}{Joan Lasenby}, \bibinfo{person}{Jure Leskovec}, \bibinfo{person}{Tie{-}Yan Liu}, \bibinfo{person}{Arjun Manrai}, \bibinfo{person}{Debora~S. Marks}, \bibinfo{person}{Bharath Ramsundar}, \bibinfo{person}{Le Song}, \bibinfo{person}{Jimeng Sun}, \bibinfo{person}{Jian Tang}, \bibinfo{person}{Petar Velickovic}, \bibinfo{person}{Max Welling}, \bibinfo{person}{Linfeng Zhang}, \bibinfo{person}{Connor~W. Coley}, \bibinfo{person}{Yoshua Bengio}, {and} \bibinfo{person}{Marinka Zitnik}.}
  \bibinfo{year}{2023}\natexlab{a}.
\newblock \showarticletitle{Scientific discovery in the age of artificial intelligence}.
\newblock \bibinfo{journal}{\emph{Nature}} \bibinfo{volume}{620}, \bibinfo{number}{7972} (\bibinfo{year}{2023}), \bibinfo{pages}{47--60}.
\newblock
\urldef\tempurl%
\url{https://doi.org/10.1038/S41586-023-06221-2}
\showDOI{\tempurl}


\bibitem[Wang et~al\mbox{.}(2024)]%
        {wang2024software}
\bibfield{author}{\bibinfo{person}{Junjie Wang}, \bibinfo{person}{Yuchao Huang}, \bibinfo{person}{Chunyang Chen}, \bibinfo{person}{Zhe Liu}, \bibinfo{person}{Song Wang}, {and} \bibinfo{person}{Qing Wang}.} \bibinfo{year}{2024}\natexlab{}.
\newblock \showarticletitle{Software Testing With Large Language Models: Survey, Landscape, and Vision}.
\newblock \bibinfo{journal}{\emph{{IEEE} Trans. Software Eng.}} \bibinfo{volume}{50}, \bibinfo{number}{4} (\bibinfo{year}{2024}), \bibinfo{pages}{911--936}.
\newblock
\urldef\tempurl%
\url{https://doi.org/10.1109/TSE.2024.3368208}
\showDOI{\tempurl}


\bibitem[Wang et~al\mbox{.}(2023b)]%
        {wang2023natural}
\bibfield{author}{\bibinfo{person}{Shangwen Wang}, \bibinfo{person}{Mingyang Geng}, \bibinfo{person}{Bo Lin}, \bibinfo{person}{Zhensu Sun}, \bibinfo{person}{Ming Wen}, \bibinfo{person}{Yepang Liu}, \bibinfo{person}{Li Li}, \bibinfo{person}{Tegawend{\'{e}}~F. Bissyand{\'{e}}}, {and} \bibinfo{person}{Xiaoguang Mao}.} \bibinfo{year}{2023}\natexlab{b}.
\newblock \showarticletitle{Natural Language to Code: How Far Are We?}. In \bibinfo{booktitle}{\emph{Proceedings of the 31st {ACM} Joint European Software Engineering Conference and Symposium on the Foundations of Software Engineering, {ESEC/FSE} 2023, San Francisco, CA, USA, December 3-9, 2023}}, \bibfield{editor}{\bibinfo{person}{Satish Chandra}, \bibinfo{person}{Kelly Blincoe}, {and} \bibinfo{person}{Paolo Tonella}} (Eds.). \bibinfo{publisher}{{ACM}}, \bibinfo{pages}{375--387}.
\newblock
\urldef\tempurl%
\url{https://doi.org/10.1145/3611643.3616323}
\showDOI{\tempurl}


\bibitem[Wei et~al\mbox{.}(2023)]%
        {wei2023towards}
\bibfield{author}{\bibinfo{person}{Xiaokai Wei}, \bibinfo{person}{Sujan~Kumar Gonugondla}, \bibinfo{person}{Shiqi Wang}, \bibinfo{person}{Wasi~Uddin Ahmad}, \bibinfo{person}{Baishakhi Ray}, \bibinfo{person}{Haifeng Qian}, \bibinfo{person}{Xiaopeng Li}, \bibinfo{person}{Varun Kumar}, \bibinfo{person}{Zijian Wang}, \bibinfo{person}{Yuchen Tian}, \bibinfo{person}{Qing Sun}, \bibinfo{person}{Ben Athiwaratkun}, \bibinfo{person}{Mingyue Shang}, \bibinfo{person}{Murali~Krishna Ramanathan}, \bibinfo{person}{Parminder Bhatia}, {and} \bibinfo{person}{Bing Xiang}.} \bibinfo{year}{2023}\natexlab{}.
\newblock \showarticletitle{Towards Greener Yet Powerful Code Generation via Quantization: An Empirical Study}. In \bibinfo{booktitle}{\emph{Proceedings of the 31st {ACM} Joint European Software Engineering Conference and Symposium on the Foundations of Software Engineering, {ESEC/FSE} 2023, San Francisco, CA, USA, December 3-9, 2023}}, \bibfield{editor}{\bibinfo{person}{Satish Chandra}, \bibinfo{person}{Kelly Blincoe}, {and} \bibinfo{person}{Paolo Tonella}} (Eds.). \bibinfo{publisher}{{ACM}}, \bibinfo{pages}{224--236}.
\newblock
\urldef\tempurl%
\url{https://doi.org/10.1145/3611643.3616302}
\showDOI{\tempurl}


\bibitem[Xia et~al\mbox{.}(2023)]%
        {xia2023automated}
\bibfield{author}{\bibinfo{person}{Chunqiu~Steven Xia}, \bibinfo{person}{Yuxiang Wei}, {and} \bibinfo{person}{Lingming Zhang}.} \bibinfo{year}{2023}\natexlab{}.
\newblock \showarticletitle{Automated Program Repair in the Era of Large Pre-trained Language Models}. In \bibinfo{booktitle}{\emph{45th {IEEE/ACM} International Conference on Software Engineering, {ICSE} 2023, Melbourne, Australia, May 14-20, 2023}}. \bibinfo{publisher}{{IEEE}}, \bibinfo{pages}{1482--1494}.
\newblock
\urldef\tempurl%
\url{https://doi.org/10.1109/ICSE48619.2023.00129}
\showDOI{\tempurl}


\bibitem[Xie et~al\mbox{.}(2024)]%
        {xie2024survey}
\bibfield{author}{\bibinfo{person}{Yutao Xie}, \bibinfo{person}{Jiayi Lin}, \bibinfo{person}{Hande Dong}, \bibinfo{person}{Lei Zhang}, {and} \bibinfo{person}{Zhonghai Wu}.} \bibinfo{year}{2024}\natexlab{}.
\newblock \showarticletitle{Survey of Code Search Based on Deep Learning}.
\newblock \bibinfo{journal}{\emph{{ACM} Trans. Softw. Eng. Methodol.}} \bibinfo{volume}{33}, \bibinfo{number}{2} (\bibinfo{year}{2024}), \bibinfo{pages}{54:1--54:42}.
\newblock
\urldef\tempurl%
\url{https://doi.org/10.1145/3628161}
\showDOI{\tempurl}


\bibitem[Yan et~al\mbox{.}(2023)]%
        {yan2023closer}
\bibfield{author}{\bibinfo{person}{Dapeng Yan}, \bibinfo{person}{Zhipeng Gao}, {and} \bibinfo{person}{Zhiming Liu}.} \bibinfo{year}{2023}\natexlab{}.
\newblock \showarticletitle{A Closer Look at Different Difficulty Levels Code Generation Abilities of ChatGPT}. In \bibinfo{booktitle}{\emph{38th {IEEE/ACM} International Conference on Automated Software Engineering, {ASE} 2023, Luxembourg, September 11-15, 2023}}. \bibinfo{publisher}{{IEEE}}, \bibinfo{pages}{1887--1898}.
\newblock
\urldef\tempurl%
\url{https://doi.org/10.1109/ASE56229.2023.00096}
\showDOI{\tempurl}


\bibitem[Yu et~al\mbox{.}(2024)]%
        {yu2024codereval}
\bibfield{author}{\bibinfo{person}{Hao Yu}, \bibinfo{person}{Bo Shen}, \bibinfo{person}{Dezhi Ran}, \bibinfo{person}{Jiaxin Zhang}, \bibinfo{person}{Qi Zhang}, \bibinfo{person}{Yuchi Ma}, \bibinfo{person}{Guangtai Liang}, \bibinfo{person}{Ying Li}, \bibinfo{person}{Qianxiang Wang}, {and} \bibinfo{person}{Tao Xie}.} \bibinfo{year}{2024}\natexlab{}.
\newblock \showarticletitle{CoderEval: {A} Benchmark of Pragmatic Code Generation with Generative Pre-trained Models}. In \bibinfo{booktitle}{\emph{Proceedings of the 46th {IEEE/ACM} International Conference on Software Engineering, {ICSE} 2024, Lisbon, Portugal, April 14-20, 2024}}. \bibinfo{publisher}{{ACM}}, \bibinfo{pages}{37:1--37:12}.
\newblock
\urldef\tempurl%
\url{https://doi.org/10.1145/3597503.3623316}
\showDOI{\tempurl}


\bibitem[Zan et~al\mbox{.}(2023)]%
        {zan2023large}
\bibfield{author}{\bibinfo{person}{Daoguang Zan}, \bibinfo{person}{Bei Chen}, \bibinfo{person}{Fengji Zhang}, \bibinfo{person}{Dianjie Lu}, \bibinfo{person}{Bingchao Wu}, \bibinfo{person}{Bei Guan}, \bibinfo{person}{Yongji Wang}, {and} \bibinfo{person}{Jian{-}Guang Lou}.} \bibinfo{year}{2023}\natexlab{}.
\newblock \showarticletitle{Large Language Models Meet NL2Code: {A} Survey}. In \bibinfo{booktitle}{\emph{Proceedings of the 61st Annual Meeting of the Association for Computational Linguistics (Volume 1: Long Papers), {ACL} 2023, Toronto, Canada, July 9-14, 2023}}, \bibfield{editor}{\bibinfo{person}{Anna Rogers}, \bibinfo{person}{Jordan~L. Boyd{-}Graber}, {and} \bibinfo{person}{Naoaki Okazaki}} (Eds.). \bibinfo{publisher}{Association for Computational Linguistics}, \bibinfo{pages}{7443--7464}.
\newblock
\urldef\tempurl%
\url{https://doi.org/10.18653/V1/2023.ACL-LONG.411}
\showDOI{\tempurl}


\bibitem[Zhang et~al\mbox{.}(2024a)]%
        {zhang2024survey}
\bibfield{author}{\bibinfo{person}{Quanjun Zhang}, \bibinfo{person}{Chunrong Fang}, \bibinfo{person}{Yuxiang Ma}, \bibinfo{person}{Weisong Sun}, {and} \bibinfo{person}{Zhenyu Chen}.} \bibinfo{year}{2024}\natexlab{a}.
\newblock \showarticletitle{A Survey of Learning-based Automated Program Repair}.
\newblock \bibinfo{journal}{\emph{{ACM} Trans. Softw. Eng. Methodol.}} \bibinfo{volume}{33}, \bibinfo{number}{2} (\bibinfo{year}{2024}), \bibinfo{pages}{55:1--55:69}.
\newblock
\urldef\tempurl%
\url{https://doi.org/10.1145/3631974}
\showDOI{\tempurl}


\bibitem[Zhang et~al\mbox{.}(2024b)]%
        {zhang2024systematic}
\bibfield{author}{\bibinfo{person}{Quanjun Zhang}, \bibinfo{person}{Chunrong Fang}, \bibinfo{person}{Yang Xie}, \bibinfo{person}{Yuxiang Ma}, \bibinfo{person}{Weisong Sun}, \bibinfo{person}{Yun Yang}, {and} \bibinfo{person}{Zhenyu Chen}.} \bibinfo{year}{2024}\natexlab{b}.
\newblock \showarticletitle{A Systematic Literature Review on Large Language Models for Automated Program Repair}.
\newblock \bibinfo{journal}{\emph{CoRR}}  \bibinfo{volume}{abs/2405.01466} (\bibinfo{year}{2024}).
\newblock
\urldef\tempurl%
\url{https://doi.org/10.48550/ARXIV.2405.01466}
\showDOI{\tempurl}
\showeprint[arXiv]{2405.01466}


\bibitem[Zhang et~al\mbox{.}(2023a)]%
        {zhang2023survey}
\bibfield{author}{\bibinfo{person}{Quanjun Zhang}, \bibinfo{person}{Chunrong Fang}, \bibinfo{person}{Yang Xie}, \bibinfo{person}{Yaxin Zhang}, \bibinfo{person}{Yun Yang}, \bibinfo{person}{Weisong Sun}, \bibinfo{person}{Shengcheng Yu}, {and} \bibinfo{person}{Zhenyu Chen}.} \bibinfo{year}{2023}\natexlab{a}.
\newblock \showarticletitle{A Survey on Large Language Models for Software Engineering}.
\newblock \bibinfo{journal}{\emph{CoRR}}  \bibinfo{volume}{abs/2312.15223} (\bibinfo{year}{2023}).
\newblock
\urldef\tempurl%
\url{https://doi.org/10.48550/ARXIV.2312.15223}
\showDOI{\tempurl}
\showeprint[arXiv]{2312.15223}


\bibitem[Zhang et~al\mbox{.}(2024c)]%
        {zhang2023pre}
\bibfield{author}{\bibinfo{person}{Quanjun Zhang}, \bibinfo{person}{Chunrong Fang}, \bibinfo{person}{Bowen Yu}, \bibinfo{person}{Weisong Sun}, \bibinfo{person}{Tongke Zhang}, {and} \bibinfo{person}{Zhenyu Chen}.} \bibinfo{year}{2024}\natexlab{c}.
\newblock \showarticletitle{Pre-Trained Model-Based Automated Software Vulnerability Repair: How Far are We?}
\newblock \bibinfo{journal}{\emph{{IEEE} Trans. Dependable Secur. Comput.}} \bibinfo{volume}{21}, \bibinfo{number}{4} (\bibinfo{year}{2024}), \bibinfo{pages}{2507--2525}.
\newblock
\urldef\tempurl%
\url{https://doi.org/10.1109/TDSC.2023.3308897}
\showDOI{\tempurl}


\bibitem[Zhang et~al\mbox{.}(2023b)]%
        {zhang2023gamma}
\bibfield{author}{\bibinfo{person}{Quanjun Zhang}, \bibinfo{person}{Chunrong Fang}, \bibinfo{person}{Tongke Zhang}, \bibinfo{person}{Bowen Yu}, \bibinfo{person}{Weisong Sun}, {and} \bibinfo{person}{Zhenyu Chen}.} \bibinfo{year}{2023}\natexlab{b}.
\newblock \showarticletitle{Gamma: Revisiting Template-Based Automated Program Repair Via Mask Prediction}. In \bibinfo{booktitle}{\emph{38th {IEEE/ACM} International Conference on Automated Software Engineering, {ASE} 2023, Luxembourg, September 11-15, 2023}}. \bibinfo{publisher}{{IEEE}}, \bibinfo{pages}{535--547}.
\newblock
\urldef\tempurl%
\url{https://doi.org/10.1109/ASE56229.2023.00063}
\showDOI{\tempurl}


\bibitem[Zhang et~al\mbox{.}(2023c)]%
        {zhang2023critical}
\bibfield{author}{\bibinfo{person}{Quanjun Zhang}, \bibinfo{person}{Tongke Zhang}, \bibinfo{person}{Juan Zhai}, \bibinfo{person}{Chunrong Fang}, \bibinfo{person}{Bowen Yu}, \bibinfo{person}{Weisong Sun}, {and} \bibinfo{person}{Zhenyu Chen}.} \bibinfo{year}{2023}\natexlab{c}.
\newblock \showarticletitle{A Critical Review of Large Language Model on Software Engineering: An Example from ChatGPT and Automated Program Repair}.
\newblock \bibinfo{journal}{\emph{CoRR}}  \bibinfo{volume}{abs/2310.08879} (\bibinfo{year}{2023}).
\newblock
\urldef\tempurl%
\url{https://doi.org/10.48550/ARXIV.2310.08879}
\showDOI{\tempurl}
\showeprint[arXiv]{2310.08879}


\bibitem[Zhang et~al\mbox{.}(2024d)]%
        {zhang2023autocoderover}
\bibfield{author}{\bibinfo{person}{Yuntong Zhang}, \bibinfo{person}{Haifeng Ruan}, \bibinfo{person}{Zhiyu Fan}, {and} \bibinfo{person}{Abhik Roychoudhury}.} \bibinfo{year}{2024}\natexlab{d}.
\newblock \showarticletitle{AutoCodeRover: Autonomous Program Improvement}.
\newblock \bibinfo{journal}{\emph{CoRR}}  \bibinfo{volume}{abs/2404.05427} (\bibinfo{year}{2024}).
\newblock
\urldef\tempurl%
\url{https://doi.org/10.48550/ARXIV.2404.05427}
\showDOI{\tempurl}
\showeprint[arXiv]{2404.05427}


\bibitem[Zhu et~al\mbox{.}(2021)]%
        {zhu2021syntax}
\bibfield{author}{\bibinfo{person}{Qihao Zhu}, \bibinfo{person}{Zeyu Sun}, \bibinfo{person}{Yuan{-}an Xiao}, \bibinfo{person}{Wenjie Zhang}, \bibinfo{person}{Kang Yuan}, \bibinfo{person}{Yingfei Xiong}, {and} \bibinfo{person}{Lu Zhang}.} \bibinfo{year}{2021}\natexlab{}.
\newblock \showarticletitle{A syntax-guided edit decoder for neural program repair}. In \bibinfo{booktitle}{\emph{{ESEC/FSE} '21: 29th {ACM} Joint European Software Engineering Conference and Symposium on the Foundations of Software Engineering, Athens, Greece, August 23-28, 2021}}, \bibfield{editor}{\bibinfo{person}{Diomidis Spinellis}, \bibinfo{person}{Georgios Gousios}, \bibinfo{person}{Marsha Chechik}, {and} \bibinfo{person}{Massimiliano~Di Penta}} (Eds.). \bibinfo{publisher}{{ACM}}, \bibinfo{pages}{341--353}.
\newblock
\urldef\tempurl%
\url{https://doi.org/10.1145/3468264.3468544}
\showDOI{\tempurl}


\bibitem[Zhu et~al\mbox{.}(2024)]%
        {zhu2024hot}
\bibfield{author}{\bibinfo{person}{Yuqi Zhu}, \bibinfo{person}{Jia Li}, \bibinfo{person}{Ge Li}, \bibinfo{person}{Yunfei Zhao}, \bibinfo{person}{Jia Li}, \bibinfo{person}{Zhi Jin}, {and} \bibinfo{person}{Hong Mei}.} \bibinfo{year}{2024}\natexlab{}.
\newblock \showarticletitle{Hot or Cold? Adaptive Temperature Sampling for Code Generation with Large Language Models}. In \bibinfo{booktitle}{\emph{Thirty-Eighth {AAAI} Conference on Artificial Intelligence, {AAAI} 2024, Thirty-Sixth Conference on Innovative Applications of Artificial Intelligence, {IAAI} 2024, Fourteenth Symposium on Educational Advances in Artificial Intelligence, {EAAI} 2014, February 20-27, 2024, Vancouver, Canada}}, \bibfield{editor}{\bibinfo{person}{Michael~J. Wooldridge}, \bibinfo{person}{Jennifer~G. Dy}, {and} \bibinfo{person}{Sriraam Natarajan}} (Eds.). \bibinfo{publisher}{{AAAI} Press}, \bibinfo{pages}{437--445}.
\newblock
\urldef\tempurl%
\url{https://doi.org/10.1609/AAAI.V38I1.27798}
\showDOI{\tempurl}


\end{thebibliography}

\end{document}